\documentclass[fleqn,usenatbib]{mnras}

\usepackage{newtxtext,newtxmath}

\usepackage[T1]{fontenc}
\usepackage{ae,aecompl}

\usepackage[dvipdfmx]{graphicx}	
\usepackage{amsmath}	
\usepackage{amssymb}	
\usepackage{indentfirst}	
\usepackage{pict2e}
\usepackage{color}
\usepackage{bm}
\usepackage{tablefootnote}
\usepackage{pict2e}

\usepackage{ulem}
\hypersetup{draft}

\title[Low-metallicity Star Cluster Formation]{
Star cluster formation and cloud dispersal by radiative feedback: \\
dependence on metallicity and compactness}

\author[Fukushima et al.]{
Hajime Fukushima$^{1}$\thanks{E-mail:fukushima@ccs.tsukuba.ac.jp},
Hidenobu Yajima$^{1}$,
Kazuyuki Sugimura$^{2}$,
Takashi Hosokawa$^{3}$,
\newauthor Kazuyuki Omukai$^{4}$ and
Tomoaki Matsumoto$^{5}$
\\
\\
$^{1}$Center for Computational Sciences, University of Tsukuba, Ten-nodai, 1-1-1 Tsukuba, Ibaraki 305-8577, Japan\\
$^{2}$ Department of Astronomy, University of Maryland, College Park, MD 20740, USA \\
$^{3}$ Department of Physics, Graduate School of Science, Kyoto University, Sakyo, Kyoto 606-8502, Japan \\
$^{4}$ Astronomical Institute, Graduate School of Science, Tohoku University, Aoba, Sendai 980-8578, Japan \\
$^{5}$ Faculty of Sustainability Studies, Hosei University, Fujimi, Chiyoda, Tokyo 102-8160, Japan
}
\date{Accepted XXX. Received YYY; in original form ZZZ}

\pubyear{2019}

\begin{document}
\label{firstpage}
\pagerange{\pageref{firstpage}--\pageref{lastpage}}
\maketitle

\begin{abstract}
We study star cluster formation in various environments with different metallicities and column densities by performing a suite of three-dimensional radiation hydrodynamics simulations. We find that the photoionization feedback from massive stars controls the star formation efficiency (SFE) in a star-forming cloud, and its impact sensitively depends on the gas metallicity $Z$ and initial cloud surface density $\Sigma$. 
At $Z=1~Z_{\odot}$, SFE increases as a power law from 0.03 at $\Sigma = 10~M_{\odot}{\rm pc^{-2}}$ to 0.3 at $\Sigma = 300~M_{\odot}{\rm pc^{-2}}$.
In low-metallicity cases $10^{-2}- 10^{-1} Z_{\odot}$, star clusters form from atomic warm gases because the molecule formation time is not short enough with respect to the cooling or dynamical time. In addition, the whole cloud is disrupted more easily by expanding H{\sc ii} bubbles which have higher temperature owing to less efficient cooling. 
With smaller dust attenuation, the ionizing radiation feedback from nearby massive stars is stronger and terminate star formation in   
dense clumps.
These effects result in inefficient star formation in low-metallicity environments: the SFE drops by a factor of $\sim 3$ at $Z=10^{-2}~Z_{\odot}$  compared to the  results for $Z=1~Z_{\odot}$, regardless of $\Sigma$.
Newborn star clusters are also gravitationally less bound. We further develop a new semi-analytical model that can reproduce the simulation results well, particularly the observed dependencies of the SFEs on the cloud surface densities and metallicities.
\end{abstract}



\section{Introduction}\label{introduction}

The formation of massive stars is an essential factor in understanding the cosmic star formation history \citep{2014ARA&A..52..415M} and galactic dynamics \citep{2018MNRAS.475..648P}. 
Stellar feedback from them regulates the total star formation rate within a galaxy \citep[e.g.,][]{2010MNRAS.402.1536S, 2012ApJ...745...50W, 2013MNRAS.428..154H, 2017ApJ...846...30Y} and induce galactic outflow \citep[e.g.,][]{1990ApJS...74..833H, 2015ApJ...809..147H}. Ionizing radiation from them destroys parent clouds and leaks out to the interstellar and intergalactic space \citep{2019ApJ...883..102K, 2020MNRAS.492.4858H}. 
Most massive stars are known to be born as members of clusters in the Milky Way \citep[e.g.][]{2008A&A...489..105S, 2008A&A...490.1071G}.
On the other hand, the origin of field massive stars in dwarf galaxies, such as Small Magellanic Cloud, is still debated \citep[e.g.,][]{2016ApJ...817..113L, 2019A&A...625A.104R}.
The formation processes of star clusters and massive stars have not been fully understood yet.

The star formation efficiency (SFE) is a pivotal quantity in the star-cluster formation as it determines whether a newborn cluster remains gravitationally bound after the cloud dispersal. As suggested by observations of nearby galaxies, the star formation in the entire galaxy proceeds slowly with the depletion timescale $\sim 1~{\rm Gyr}$ \citep[e.g.][]{1998ApJ...498..541K,2008AJ....136.2846B,2013ApJ...772L..13M}.
On the other hand, the star formation in an individual giant molecular cloud (GMC) only lasts for $\sim 10~{\rm Myr}$ \citep{2001ApJ...562..852H, 2010ARA&A..48..547F} by considering observed numbers of GMCs in local galaxies \citep[e.g.,][]{2020MNRAS.493.2872C}.
This implies that only a small fraction of the molecular gas is converted into stars and the rest returns into an interstellar or intergalactic medium. 
Recently, \citet{Kruijssen_Nature_2019} showed that the lifetime of GMCs are about one dynamical time, and the SFE is only a few percents, by analyzing the spatial de-correlation between young stars and star-forming regions in the spiral galaxy NGC300.
They indicated that GMCs are disrupted before the end of the lifetimes of OB stars, and thus pre-supernova feedback, such as ultraviolet (UV) radiation feedback or stellar winds, regulated star formation in GMCs.

Extreme ultraviolet (EUV; $13.6~{\rm eV} \leq h \nu \lesssim 1~{\rm keV}$) radiation from massive stars creates H{\sc ii} bubbles and heat up the gas to $\sim 10^{4}~{\rm K}$.
The H{\sc ii} bubbles expand rapidly because of the high thermal pressure so that heated materials are photoevaporated against the gravitational pull from the gas and stars.
Based on simple analytical models, previous studies suggested that the radiative feedback suppresses star formation in GMCs, resulting in the low SFEs \citep[e.g.,][]{1997ApJ...476..166W, 2002ApJ...566..302M, 2006ApJ...653..361K, 2016ApJ...819..137K}. 
\citet{2010ApJ...710L.142F} derived an analytical expression for the SFE as a function of the surface density of the clouds. 
In reality, however, the supersonic turbulent motion shapes hierarchical cloud structures. H{\sc ii} regions forming around newborn clusters expand across such complex structures and finally destroy the GMCs.

Numerical simulations have been utilized \citep[e.g.,][]{2010ApJ...715.1302V, 2012MNRAS.427.2852D} to investigate the star cluster formation and cloud dispersal through the realistic evolution.
\citet{2017MNRAS.471.4844G} showed the disruption of molecular clouds due to the radiative feedback by using radiation hydrodynamics (RHD) simulations with a diffuse approximation for the radiative transfer \citep[{\sc RAMSES-RT}:][]{2013MNRAS.436.2188R}. 
They showed that the radiative feedback quenches the star formation immediately, and the SFE increases with increasing cloud column density. Using the same simulation code, \citet{2019MNRAS.489.1880H} investigated this problem in a wider parameter range of initial gas clouds and showed that the massive-end slope of simulated sink mass function was similar to that of Salpeter-like initial mass function. 

Recently, \citet{2018ApJ...859...68K} showed the cloud dispersal occurred within 2-10 Myr after the onset of radiative feedback using the RHD simulations with a ray-tracing scheme. 
They found a close relationship between the SFEs and the surface densities of gas clouds. 
However, their simulations were limited to the clouds with metallicity $Z=1~Z_{\odot}$. Varying the metallicity potentially changes the star formation and feedback processes. For instance, a smaller amount of dust grains results in less attenuation of the stellar UV radiation, which strengthens the feedback effect \citep[e.g.,][]{2020arXiv200402364F}.

In this paper, we study the star-cluster formation and resulting cloud dispersal for various cases with metallicities $Z=10^{-2} - 1~ Z_{\odot}$. 
We perform three-dimensional RHD simulations with {\sc SFUMATO-RT}  
\citep{2020ApJ...892L..14S}, an Eulerian adaptive mesh refinement (AMR) hydrodynamics code \citep{2007PASJ...59..905M} coupled with the radiative transfer solver using the ray-tracing method.
Also, we calculate the gas temperature considering metal and dust cooling, while \citet{2018ApJ...859...68K} assumed the temperature with an approximation based on local ionization degree. 
The temperature of ionized gas decreases as the metallicity increases due to the metal-line cooling, e.g., O{\sc ii} and O{\sc iii} \citep{2006agna.book.....O}. As discussed in \citet{2019MNRAS.489.1880H}, an H{\sc ii} bubble expands earlier at lower metallicities, resulting in the shorter lifetime of the clouds.

In the current work, we consistently solve the thermal and chemical states of the neutral and molecular gases, from which stars form. With $Z \gtrsim 10^{-2}~Z_{\odot}$, CO molecules ($n_{\rm H} < 10^{5}~{\rm cm^{-3}}$) and dust grains ($n_{\rm H} > 10^{5}~{\rm cm^{-3}}$) are efficient coolant for the molecular gas \citep[e.g.,][]{2005ApJ...626..627O}, with which the temperature could drop to the CMB value. The atomic gas also cools mainly via O{\sc i} and C{\sc ii} fine-structure line emission. Far-ultraviolet (FUV; $6.0 ~ {\rm eV} \leq h \nu \leq 13.6 ~ {\rm eV}$) photons are the dominant heating sources for the neutral gas via the photodissociation of hydrogen molecules and photoelectric heating of dust grains. As a result, the H{\sc ii} bubbles are surrounded by a warm neutral layer where the temperature is raised up to a few $100~{\rm K}$ in the case of $Z = Z_\odot$ \citep[so-called the photodissociation region or PDR; e.g.,][]{HT99,2005ApJ...623..917H, 2006ApJ...646..240H}. Our simulations show that at very low metallicity of $Z=10^{-2}~Z_\odot$ the whole star-forming clouds are mostly dominated by such warm atomic gases, and that massive star clusters directly form from the atomic clouds. 

We organize the rest of the paper as follows. 
In Section \ref{numerical_method}, we describe the numerical method and setup of our simulations. We present our simulation results in Section \ref{Results}.
We then develop a semi-analytical model that helps us to interpret numerical results in Section \ref{semi_analytical_model}. Section \ref{summary} is for summary and discussions. We describe the details of the numerical simulations and the fitting formula of the fraction of hydrogen ionization in an H{\sc ii} region in Appendix \ref{apdA} and \ref{fion}.

\section{Numerical Method} \label{numerical_method}

We perform RHD simulations with SFUMATO-RT \citep{2020ApJ...892L..14S}, a modified version of the self-gravitational magnetohydrodynamics code with AMR, SFUMATO \citep{2007PASJ...59..905M,2015ApJ...801...77M}.
SFUMATO-RT has been developed to solve radiative transfer with the adaptive ray-tracing method \citep{2002MNRAS.330L..53A} and non-equilibrium chemistry of the primordial gas.
For the current purpose, we further add the effects of heavy elements (e.g., dust grains, atomic carbon, oxygen, etc.) as described in Section~\ref{chemical_thermal}.

The hydrodynamics is solved with the Cartesian coordinates.
We initially set uniform 64 meshes in each direction and refine them during the simulations as the gas density gradually increases according to the refinement criterion of at least 5 meshes for one Jeans length.
The computational box at the coarsest grid level is large enough to cover the whole cloud, $-2R_{\rm cl} \leqq x \leqq 2R_{\rm cl}$ on a side, where $R_{\rm cl}$ is the cloud radius and the origin of coordinates ($x, y, z$) is at the box center.
We set the maximum level of the refinement so that the maximum resolution becomes $0.078~{\rm pc}$ regardless of the box size.
In the case with the maximum level $l_{\rm max} = 5$, the effective resolution corresponds to $64 \times 2 ^5 = 2048$ uniform meshes on a side.
We solve the following set of equations for compressible hydrodynamics:
\begin{eqnarray}
	\frac{\partial \rho}{\partial t} + \nabla \cdot \left( \rho \bm{v} \right) = 0, \label{basic_eq1}
\end{eqnarray}
\begin{eqnarray}
	\frac{\partial \left( \rho \bm{v} \right)}{\partial t} + \nabla \cdot \left( \rho \bm{v} \otimes \bm{v} \right) + \nabla P  =  \rho \left( \bm{g} + \bm{f} \right) , \label{basic_eq2}
\end{eqnarray}
\begin{eqnarray}
	\frac{\partial \left( \rho E \right)}{\partial t} + \nabla \cdot \left[ \left( \rho E + P \right) \bf{v} \right] = \rho \left( \bf{g} + \bf{f} \right) \cdot \bf{v} + \Gamma - \Lambda, \label{basic_eq3}
\end{eqnarray}
\begin{eqnarray}
	E = \frac{| {\bf v} |^2}{2} + \left( \gamma - 1 \right)^{-1} \frac{P}{\rho}, \label{basic_eq4}
\end{eqnarray}
where $\rho$, $P$, $\bf{v}$, $\bf{g}$, $E$, $\Gamma$ and $\Lambda$ are the density, pressure, velocity, gravitational force, total energy, the heating and cooling functions.
In Equation \eqref{basic_eq2} and \eqref{basic_eq3}, $\bf{f}$ is the radiation forces per unit mass.
We consider the absorption of ionizing photons by H{\sc i} and dust grains, and that of non-ionizing photons by dust grains.
We also calculate the specific heat $\gamma$ as a function of temperature and chemical abundances \citep{1998ApJ...508..141O}.  
In this study, we include thermal processes due to heavy elements in addition to those in the primordial gas.

\subsection{Chemical and thermal processes}\label{chemical_thermal}

We update the chemistry module of SFUMATO-RT to calculate the chemical and thermal evolution of the gas.
We consider the chemical network of 11 species: $\rm H$, $\rm H_2$, $\rm H^{-}$, $\rm H^{+}$, $\rm H_{2}^{+}$, $\rm e$, $\rm CO$, C{\sc ii}, O{\sc i}, O{\sc ii} and O{\sc iii}.
To follow the CO formation, we adopt the simple chemical network of \citet{1997ApJ...482..796N}.
In this model, we only explicitly include C{\sc ii} and CO as carbon species. 
CO is assumed to form via hydrocarbon radicals without C{\sc i} in the chemical network. \citet{2012MNRAS.421..116G} and \citet{2017ApJ...843...38G} showed limitations of such a simplified treatment and provided the alternative extended network. However, our qualitative results of the cluster formation and cloud dispersal do not rely on those details because C{\sc ii} becomes the dominant coolant instead of CO molecules \citep[e.g.,][]{2012MNRAS.421....9G}.
As in \citet{2020arXiv200402364F}, we assume that the neutral fraction of oxygen is the same as hydrogen and that O{\sc ii} and O{\sc iii} are in the chemical equilibrium with photoionization and recombination.
The metal-line cooling by O{\sc ii} and O{\sc iii} becomes the dominant coolant in H{\sc ii} regions
at metallicity as high as $Z=1~Z_{\odot}$. 

The total heating/cooling rates in Equation \eqref{basic_eq3} are given as 
\begin{eqnarray}
 \Gamma = \Gamma_{\rm prim} + \Gamma_{\rm H_2, d} + \Gamma_{\rm pe}, \label{heating_func}
\end{eqnarray}
\begin{eqnarray}
\Lambda = \Lambda_{\rm prim} + \Lambda_{\rm d} + \Lambda_{\rm m, line}, \label{cooling_func}
\end{eqnarray}
where $\Gamma_{\rm prim}$ and $\Lambda_{\rm prim}$ are the thermal processes related to primordial star formation as in \citet{2016ApJ...824..119H}.
In this study, we newly add heating of $\rm H_{\rm 2}$ formation on dust grains $(\Gamma_{\rm H_2, d})$ and energy transport between dust grains and gas $(\Lambda_{\rm d})$ as in \citet{2020arXiv200402364F}.
In Equation \eqref{heating_func}, $\Gamma_{\rm pe}$ is the photoelectric heating rate, and we describe its details in Appendix \ref{apdA}.
We consider C{\sc ii}, CO, O{\sc i}, O{\sc ii} and O{\sc iii} as the metal-line cooling $(\Lambda_{\rm m, line})$.
To calculate the cooling rates of C{\sc ii}, O{\sc i}, O{\sc ii}, and O{\sc iii}, we solve the statistical equilibrium of each energy level as in \citet{2020arXiv200402364F}.
For the cooling rate of CO rotational transitions, we use the fitting function by \citet{2010ApJ...722.1793O}.
We assume that the dust-to-gas mass ratio is 0.01 at $Z=1~Z_{\odot}$ and decreases linearly in proportion to the metallicity.
We adopt the MRN mixture \citep{1977ApJ...217..425M} for the grain size distribution.

The radiative processes considered in our simulations are photoionization of H{\sc i} and O{\sc ii}, photodissociation of $\rm H_2$ and CO, and photoelectric heating of dust grains.
We allow that the gas and dust have different temperatures due to the ineffective coupling. 
The dust temperature is calculated from the energy balance between the energy exchange with gas and absorption/emission of radiation.
Stellar irradiation heats dust grains up, and the sublimation distance locates at $\sim 100~{\rm au}$ from massive stars \citep{2018MNRAS.473.4754F}.
Such small structure is represented by the sink regions and the sublimation distance effectively corresponds to the sink radius in our simulations. In these cases, we may overestimate the strength of radiation pressure caused by stellar irradiation \citep{2018MNRAS.480.3468K}.
Nonetheless, the radiation pressure effect does not contribute much to the cloud dispersal in the cases considered below (see Sec. \ref{radiative_feedback_on_GMCs}), and our results are not affected by this approximation. 
We also consider the radiation pressure via the direct stellar photons ($\bf{f}$ in Eq. \ref{basic_eq2} and \ref{basic_eq3}).
The optical depth of a star-forming cloud for dust thermal radiation is estimated as
\begin{eqnarray}
	\tau_{\rm IR} = \rho R_{\rm cl} \kappa_{\rm IR} = 6.3 \times 10^{-2} \left( \frac{\Sigma}{80~M_{\odot}{\rm pc^{-2}}} \right) \left( \frac{Z}{Z_{\odot}} \right), \label{optical_depth}
\end{eqnarray}
where $R_{\rm cl}$ and $\Sigma$ are the radius and surface density of the cloud.
Here, we use the opacity for thermal emission as $\kappa_{\rm IR} = 5 ~ {\rm cm^{2} \, g^{-1}} \left( Z / Z_{\odot}\right)$.
Star-forming clouds examined in this study are optically thin for dust thermal emission. Thus we ignore the radiation pressure of dust thermal emission, although it can be significant in clouds more compact or metal-enriched than those examined in this paper \citep[e.g.,][]{2010ApJ...709..191M, 2015ApJ...809..187S}.
We describe the method of radiative transfer in more detail in Appendix \ref{apdA}.

We set the temperature floor at $T=10~{\rm K}$ in our simulations, supposing the CMB temperature at the redshift $z \simeq 2.5$. 
This floor does not affect our results because the gas temperature is normally much higher than the floor value, particularly in low-metallicity cases. 
We do not consider stellar UV background radiation, which would permeate into the cloud externally. 
We discuss this effect later in Section \ref{Rn_in_starforming_region}. 

\subsection{Sink particles and radiation sources}\label{radsource}

\begin{table*}
 	\caption{Models considered}
 	\label{tab1}
 	\centering

\begin{tabular}{|l|c|c|c|c|c|c|c|c|c|} \hline \hline
model & $M_{\rm cl} \, [ \, M_{\odot} \, ]$ & $R_{\rm cl}\, [ \, {\rm pc} \, ]$ & $Z  \, [ \, Z_{\odot} \, ] $ & $n_{\rm ini} \, [ \, {\rm cm^{-3}} \, ]$ & $\Sigma \, [\, M_{\odot} \, {\rm pc^{-2}} \, ]$ & $\sigma_0 \, [ \, {\rm km s^{-1}} \, ]$  & $t_{\rm ff} \, [ \, {\rm Myr} \, ]$ & $l_{\rm max}$ & Feedback  \\ \hline
M4R10Z0 & $10^{4}$ & $10$ & $1$ & $70$ & $32$ & $1.1$ & 5.2 & 3 & PI${}^\text{a}$, RP${}^\text{b}$  \\
M4R20Z0 & $10^{4}$ & $20$ & $1$ & $8.7$ & $8$ & $0.80$   & 15 & 4 & PI, RP \\
M5R10Z0 & $10^{5}$ & $10$ & $1$ & $700$ & $320$ & $3.6$ & 1.7& 3 & PI, RP \\
M5R20Z0 & $10^{5}$ & $20$ & $1$ & $87$ & $80$ & $2.5$  & 4.7 & 4  & PI, RP \\
M5R40Z0 & $10^{5}$ & $40$ & $1$ & $11$ & $20$ & $1.8$  & 13  & 5 & PI, RP \\
M4R10Z-1 & $10^{4}$ & $10$ & $10^{-1}$ & $70$  & $32$ & $1.1$ & 5.2 & 3& PI, RP  \\
M4R20Z-1& $10^{4}$ & $20$  & $10^{-1}$ & $8.7$  & $8$ & $0.80$   & 15& 4 & PI, RP \\
M5R10Z-1 & $10^{5}$ & $10$ & $10^{-1}$ & $700$  & $320$ & $3.6$  & 1.7  & 3 & PI, RP \\
M5R20Z-1 & $10^{5}$ & $20$ & $10^{-1}$ & $87$ & $80$ & $2.5$ & 4.7  & 4 & PI, RP \\
M5R40Z-1 & $10^{5}$ & $40$ & $10^{-1}$ & $11$  & $20$ & $1.8$   & 13  & 5& PI, RP  \\
M4R10Z-2 & $10^{4}$ & $10$ & $10^{-2}$ & $70$ & $32$  & $1.1$ & 5.2 & 3 & PI, RP \\
M4R20Z-2& $10^{4}$ & $20$  & $10^{-2}$ & $8.7$   & $8$ & $0.80$    & 15  & 4& PI, RP  \\
M5R10Z-2 & $10^{5}$ & $10$ & $10^{-2}$ & $700$ & $320$ & $3.6$   & 1.7  & 3 & PI, RP \\
M5R20Z-2 & $10^{5}$ & $20$ & $10^{-2}$ & $87$ & $80$ & $2.5$  & 4.7  & 4 & PI, RP \\
M5R40Z-2 & $10^{5}$ & $40$ & $10^{-2}$ & $11$ & $20$ & $1.8$  & 13   & 5& PI, RP  \\
M5R20Z0PI & $10^{5}$ & $20$ & $1$ & $87$ & $80$ & $2.5$  & 4.7 & 4  & PI \\
M5R20Z0RP & $10^{5}$ & $20$ & $1$ & $87$ & $80$ & $2.5$  & 4.7 & 4  & RP \\
\hline 
\end{tabular}
     \begin{minipage}{1 \hsize}
      Notes. Column 1: model names, Column 2: cloud masses, Column 3: cloud radii, Column 4: metallicities, Column5: initial number densities, Column 6: surface densities, Column 7: three-dimensional velocity dispersions, Column 8: free fall times, Column 9: maximum refinement levels 
      
      ${}^\text{a}$"PI" represents photoionization feedback
      
      ${}^\text{b}$"RP" represents radiation-pressure feedback
      
    \end{minipage}
\end{table*}


We employ the sink particle technique to follow long-term evolution without resolving very dense structures by adding finer grid refinement levels.
We insert a sink particle in a mesh that satisfies the following conditions \citep{2010ApJ...713..269F, 2015ApJ...801...77M}: 1) the gas density is above the threshold value, 2) the gravitational potential is a local minimum, 3) the velocity divergence $\nabla \cdot \bm{v}$ and all the eigenvalues of the velocity gradient tensor $\nabla \bm{v}$ are negative and 4) the sum of the thermal, kinetic and gravitational energies inside the sink radius is negative ($E_{\rm th} + E_{\rm kin} + E_{\rm gr} < 0$).
We determine the density threshold in a way proposed by \citet{2013ApJS..204....8G}, as in \citet{2018ApJ...859...68K}.
Assuming the density profile of Larson-Penston solution \citep{1969MNRAS.145..271L, 1969MNRAS.144..425P} around the density peak, we set the density threshold as $\rho_{\rm thr} = 8.86 c_{\rm s}^2/(\pi G \Delta x^2)=5.34\times10^{-19}~{\rm g \, cm^{-3}}$, where $\Delta x$ is the minimum mesh size ($\Delta x = 0.078~{\rm pc}$) and $c_{\rm s}$ is the sound speed of gas with temperature $T=20~{\rm K}$.
In practice, if the gas temperature is higher than 20K, the condition (4) of the sink creation is not satisfied at $\rho_{\rm thr}$ and the gas density becomes higher than $\rho_{\rm thr}$ until all of the conditions are fulfilled.
Therefore, the threshold density $\rho_{\rm thr}$ should be regarded as the lower limit for the sink creation.
We set the sink radius $r_{\rm sink} = 2 \Delta x$ and merge nearby sink particles if their separation distances are smaller than $2r_{\rm sink}$.
The velocity of the merged sink particles is determined from momentum conservation.

We regard each sink particle as a star cluster, as we cannot resolve individual stars due to high computational cost, and calculate the transfer of radiation emitted from each cluster particle. To evaluate the luminosity and spectrum of the star cluster, we take the average of the stellar isochrone given by \citet{2015MNRAS.452.1068C}, where the Chabrier IMF \citep{2003PASP..115..763C} with a stellar-mass range from $0.1~M_{\odot}$ to $150~M_{\odot}$ is used.
In our simulations, the duration time of star formation is on the order of Myr, and thus we use their isochrone at $t=1~{\rm Myr}$. We assume that low-mass cluster particles with $< 50~M_{\odot}$ do not emit ionizing photons, neglecting the weak contributions from the smaller clusters. This is a reasonable approximation as the expected number of massive stars $(\gtrsim 10~M_{\odot})$ is less than unity for such small clusters. With this procedure, we significantly reduce the computational costs of solving radiation transfer.

\subsection{Initial conditions}

Table \ref{tab1} summarizes the models considered in this paper.
We set uniform density spheres with turbulent velocities as the initial conditions.
We consider five different clouds with masses $M_{\rm cl} = 10^{4}$ and $10^{5}~M_{\odot}$ and size $R_{\rm cl}=10\,\text{--}\,40~{\rm pc}$, which are consistent with the observed properties of galactic and extragalactic GMCs \citep[e.g.,][]{2018ApJ...857...19F}.
In Table \ref{tab1}, the surface density is defined as $\Sigma = M_{\rm cl}/ (\pi R_{\rm cl}^2)$.
The gas metallicity is assumed to be the solar value in the fiducial model, but we also consider lower-metallicity environments with $10^{-2}$ and $10^{-1}~Z_{\odot}$.
To see the impact of the individual feedback effect, we also simulate the cases where we only consider either photoionization (M5R20Z0PI) or radiation pressure induced by dust grains (M5R20Z0RP) for a star-forming cloud with $(M_{\rm cl}, R_{\rm cl}, Z)=(10^{5}M_{\odot}, 20{\rm pc}, 1Z_{\odot})$.
We set the initial gas temperature of $20~{\rm K}$ irrespective of metallicity, which is the approximate value of collapsing clouds when the central density reaches $n_{\rm ini}\sim 10 \,\text{--}\,10^3 ~{\rm cm^{-3}}$ \citep{2005ApJ...626..627O}, the initial values in our simulations.
Note, however, that the initial turbulent velocity is supersonic, as we will describe below, and thus the dynamics of the clouds are almost independent of the initial gas temperature. 
We also assume the atomic gas at the initial state. We set the abundance of hydrogen molecules is $y_{\rm H_2} = 10^{-3}$ \footnote{The abundance of species $i$ is given as $y_{ i}$ that is the ratio of the number density of the species i to that of atomic hydrogen.}, and other species are in atomic states. This assumption is motivated by theoretical prediction for low-metallicity star-forming regions \citep{2012ApJ...759....9K}. On the other hand, observations show that the star-forming clouds are almost fully molecular at $Z=1~Z_{\odot}$. In our simulations, molecules rapidly form in high-density parts within the clouds before the first cluster particle appears. The star formation then mostly occurs in the molecular clouds surrounded by rarefied atomic envelope. We also show that this is not necessarily the case for lower metallicity cases, where star clusters are formed in almost atomic clouds.

The supersonic turbulent velocity fields are added at the beginning of the simulations.
As in the previous studies \citep[e.g.,][]{2015ApJ...801...77M, 2018ApJ...859...68K}, we assume the velocity power spectrum of $P(k) \propto k^{-4}$ with wavenumber $k$, which corresponds to the velocity dispersion law of $\sigma(l) \propto l^{1/2}$ with length scale $l$ \citep{1981MNRAS.194..809L}.
The amplitude is given by setting the virial parameter
\begin{eqnarray}
	\alpha_{0} = \frac{5 \sigma_0^2 R_{\rm cl}}{3 G M_{\rm cl}} \label{1.1}
\end{eqnarray}
to be 0.5, 
where $\sigma_0$ is the three-dimensional velocity dispersion.
We impose the same random turbulent fields in all models.
The strength of the turbulent motions used in our simulations is comparable to that considered in \citet{2019MNRAS.489.1880H} ($\alpha= 0.4$). Note that \citet{2018ApJ...859...68K} adopted a larger value of $\alpha=2.0$, i.e., the clouds less gravitationally unstable with stronger turbulence support. 

\section{Results}\label{Results}

We describe the results of our simulations in this section.
In Section \ref{fiducial_case}, we show the time evolution of the cloud being destroyed by the radiative feedback from a newborn star cluster, first for the fiducial solar-metallicity model and then for the low-metallicity model with $Z=10^{-2}~Z_{\odot}$ (models M5R20Z0 and M5R20Z-2 in Table \ref{tab1}). 
In Section \ref{propGMC}, we present the dependences of the SFEs and the timescale until star formation ceases on the cloud surface density and metallicity, obtained from all the examined cases. In Section \ref{properties_of_star_clusters}, we show the time evolution of the half-mass radius and the mass distribution of star cluster particles to investigate the metallicity dependence of the gravitational boundness of the newborn clusters.

\subsection{Star cluster formation and cloud dispersal} \label{fiducial_case}
\subsubsection{Fiducial solar-metallicity model}\label{1Zsun}
 \begin{figure*}
 \begin{center}
 \includegraphics[width=170mm]{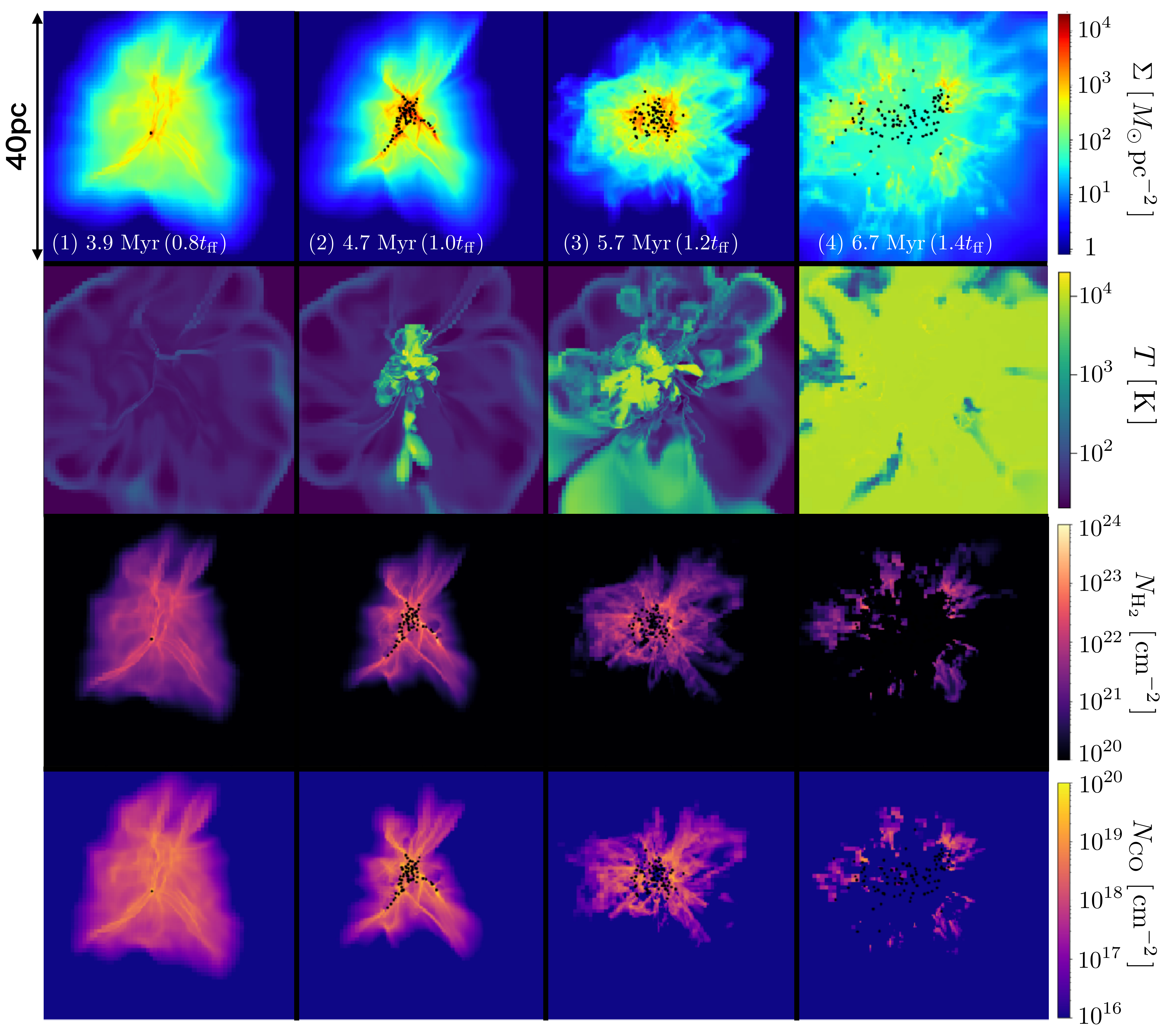}
 \end{center}
 \caption{The cloud dispersal and formation of the star cluster for the fiducial model with $(M_{\rm cl}, R_{\rm cl}, Z) = (10^5 ~M_{\odot}, 20~{\rm pc}, 1 Z_{\odot})$.
 We plot the surface density, temperature (on a slice) and the column densities of $\rm H_{2}$ and $\rm CO$ molecules from top to bottom. The four panels at each row show the snapshots at the epochs of $t=3.9, 4.7, 5.7$ and $6.3 ~{\rm Myr}$. The black dots represent the positions of the star cluster particles.
  }
 \label{zu1}
 \end{figure*}
 \begin{figure*}
 \begin{center}
 \includegraphics[width=130mm]{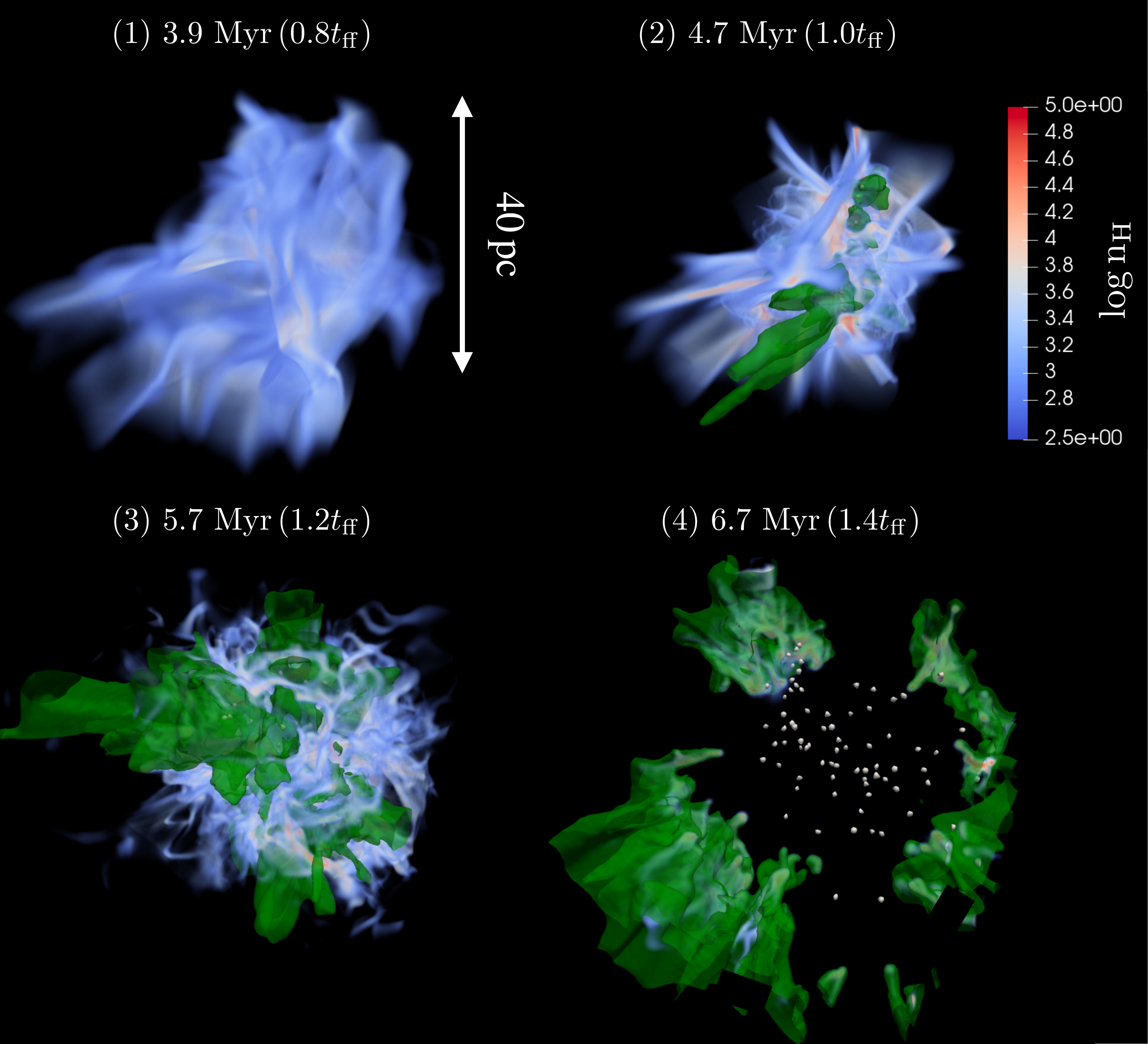}
 \end{center}
 \caption{  
 3D volume rendering of the density field, together with the ionization fronts (green surface) where the ionization degree is $0.8$, for the same case as in Figure \ref{zu1} with $(M_{\rm cl}, R_{\rm cl}, Z) = (10^5M_{\odot}, 20{\rm pc}, 1Z_{\odot})$.
 Each panel shows the snapshot at the same epochs as in Figure \ref{zu1}. The white dots in the panel (4) represent the positions of the star cluster particles. The box size is $40~{\rm pc}$ on a side.
  }
 \label{zu200108}
 \end{figure*}
 
We first present the case with $(M_{\rm cl}, R_{\rm cl}, Z) = (10^{5}M_{\odot}, 20{\rm pc}, 1Z_{\odot})$ as the fiducial model, which has the moderate initial cloud surface density $\Sigma \simeq 80~M_\odot~{\rm pc}^{-2}$ (see M5R20Z0 in Table \ref{tab1}).
Figure \ref{zu1} shows the evolution of the surface density $\Sigma$, temperature $T$ on the $xy$-plane of $z=0$, the column densities of $\rm H_{2}$ ($N_{\rm H_2}$) and $\rm CO$ ($N_{\rm CO}$).
In Figure \ref{zu200108}, we show the volume rendering of the number density, together with the positions of star cluster particles and ionization front at the same snapshots as in Figure \ref{zu1}.

Early on, the turbulent motions and self-gravity control the evolution of the cloud.
Sheet-like structures are induced by turbulent motions.
At the intersection of these sheets, filamentary density structures appear.
Then, the gas density gradually increases along the filaments due to self-gravity.
At this time, the gas temperature is generally less than $30~{\rm K}$ because of the cooling by heavy elements (C{\sc ii}, O{\sc i}, and CO).
The first star-cluster particle forms in the central region of the cloud when the elapsed time of the simulation is $3.8~{\rm Myr}$ (Fig\ref{zu1}-1, \ref{zu200108}-1).
After that, stars continuously form along the filaments in the central 10 pc region of the cloud, as shown in the snapshot at $4.7~{\rm Myr}$, which is equal to one free-fall time of the cloud, $t_{\rm ff}$ (Fig\ref{zu1}-2, \ref{zu200108}-2).
Ionizing photons emitted by the stars begin to form H{\sc ii} regions, which expand outward
due to enhanced thermal pressure of the ionized gas with $8000 - 10^4 ~{\rm K}$.
The ionization fronts propagate in gap regions between the filament structures since H{\sc ii} regions expand more quickly at lower densities.
The filamentary structures are not disrupted by photoionization, but the expansion of H{\sc ii} regions starts to push them out ($5.7~{\rm Myr}$, Fig\ref{zu1}-3, \ref{zu200108}-3).
Finally, most of them are swept out and the star formation is quenched at $6.7~{\rm Myr}$ (Fig\ref{zu1}-4, \ref{zu200108}-4).
Even though the total emissivity of ionizing photons is large enough to completely ionize all the gas in the initial homogeneous cloud,
H{\sc ii} regions cannot spread to the entire domains, and some filamentary structures survive until the final time of the simulation.

As the collapse of the cloud proceeds, $\rm H_2$ and CO molecules gradually form in dense regions.
In the snapshots at $3.9~{\rm Myr}$ and $4.7~{\rm Myr}$ of Figure \ref{zu1}, $\rm H_2$ and CO molecules form in the entire region of the cloud.
These molecules, however, cannot survive except in dense structures after FUV photons from stars begin to photodissociate them.
Therefore, the column densities of $\rm H_2$ and CO molecules only trace the dense filamentary structures in the snapshots at $5.7~{\rm Myr}$ and $6.7~{\rm Myr}$ of Figure \ref{zu1}.

 \begin{figure}
 \begin{center}
 \includegraphics[width=\columnwidth]{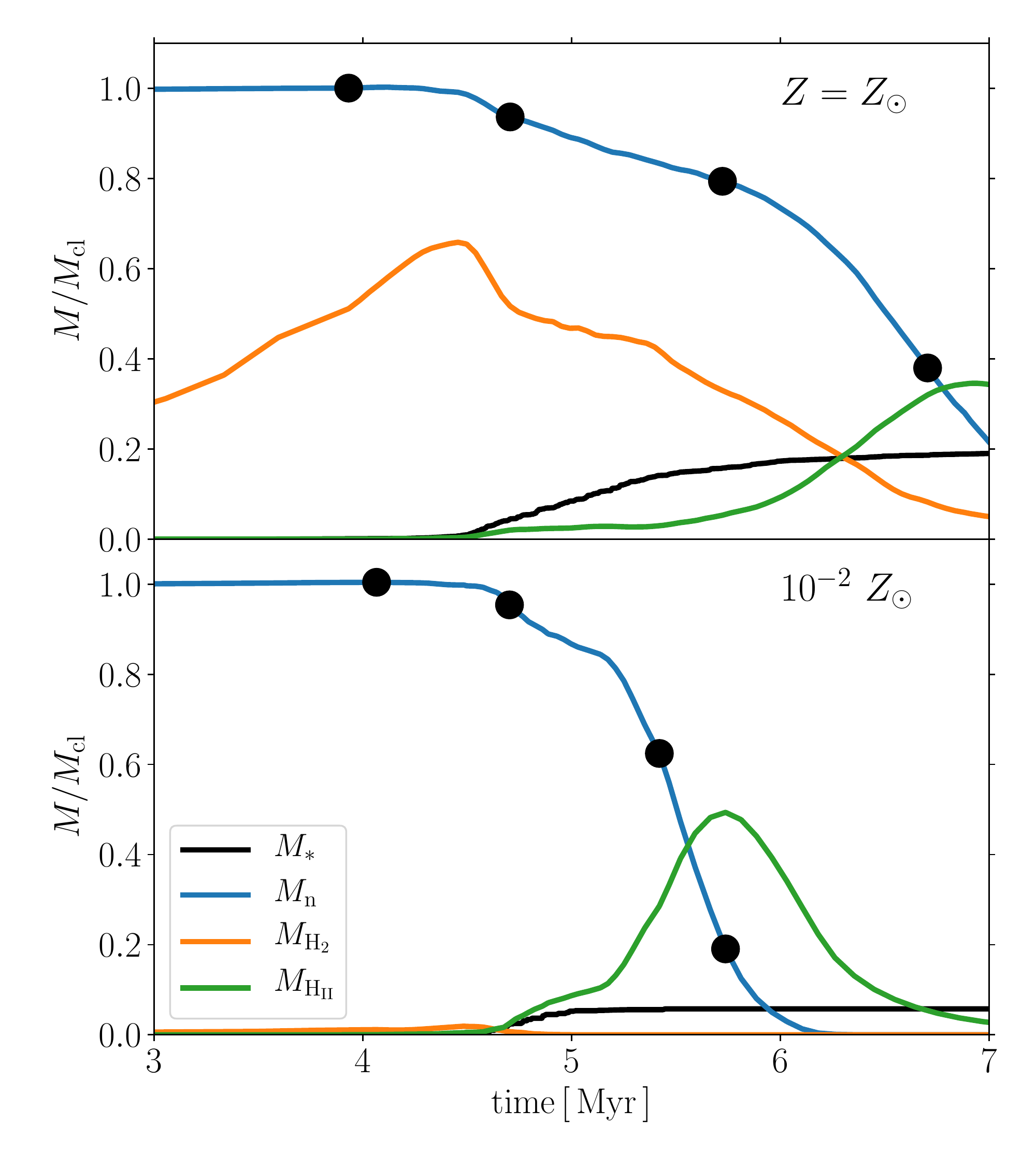}
 \end{center}
 \caption{ 
The time evolution of the mass of each component in the cluster-forming clouds with mass $M_{\rm cl} = 10^{5}M_{\odot}$, radius $R_{\rm cl} = 20{\rm pc}$, and metallicity $Z=1 Z_{\odot}$ (top) and $10^{-2}Z_{\odot}$ (bottom). 
Each line represents the total stellar mass $M_*$ (black), neutral gas mass $M_{\rm n}$ (blue), $\rm H_2$ molecule mass (orange) and ionized gas mass $M_{\rm HII}$ (green).
The filled circles mark the four epochs for which we show the snapshots in Figures \ref{zu1} and \ref{zu200108}.
 }
 \label{zu1.2}
 \end{figure}
 
The top panel of Figure \ref{zu1.2} shows the time evolution of the mass in each component, i.e., the stars, neutral, molecular and ionized gases.
Here, the neutral gas is defined as the neutral fraction of hydrogen is higher than 0.9.
Hydrogen molecules form via the dust surface reaction, which increases the amount of $\rm H_2$ to 0.6 times the cloud mass at $4.5~{\rm Myr}$.
After that, the stellar mass grows while the $\rm H_2$ mass decreases due to the photodissociation by stellar FUV radiation.
The ionized gas emerges as soon as the star formation begins, but its mass is kept small until it rapidly increases at $6.5~{\rm Myr}$ as 
the H{\sc ii} regions expand and quench the star formation.
At the final step of the simulation, clumps that contain the residual neutral component are being  ionized gradually, without forming stars anymore within them.

In the above case, star formation starts at $3.8~{\rm Myr}$ and is quenched at $6.7 ~ {\rm Myr}$.
Hence, the duration of the star formation ($\sim 3~{\rm Myr}$) is less than the free-fall time of the cloud $(\sim 4.7{\rm Myr})$.
The SFE defined by the final stellar mass normalized by the initial cloud mass is 19 percent.
\citet{2018ApJ...859...68K} adopted a similar setup although they use $\alpha = 2.0$ (see Eq. \ref{1.1}) as the initial condition. The stronger turbulent field suppresses the star formation more effectively, but the resulting SFE is in the same order of $10$ percent. 
On the other hand, \citet{2019MNRAS.489.1880H} adopted centrally concentrated density profiles and considered magnetic fields.
Nevertheless, the SFEs in their simulations have similar values to ours.
In our simulations, the clouds collapse in a self-similar fashion \citep{1969MNRAS.145..271L} in the early stages because the turbulent motion cannot support the collapsing clouds with $\alpha = 0.5$.
When the star formation takes place, the density structure resembles the power-law profile with the high-density core at the centre as in the initial setup of \citet{2019MNRAS.489.1880H}.
Besides, they adopted very similar strength of turbulent motions to our simulations ($\alpha = 0.4$) and magnetic fields weaker than the turbulence. 
Therefore, the small difference in the SFEs might be attributable to the similar virial parameter $\alpha$.

\subsubsection{Low-metallicity model with $Z = 10^{-2}~Z_\odot$}\label{low_metallicity_cloud}
 \begin{figure*}
 \begin{center}
 \includegraphics[width=170mm]{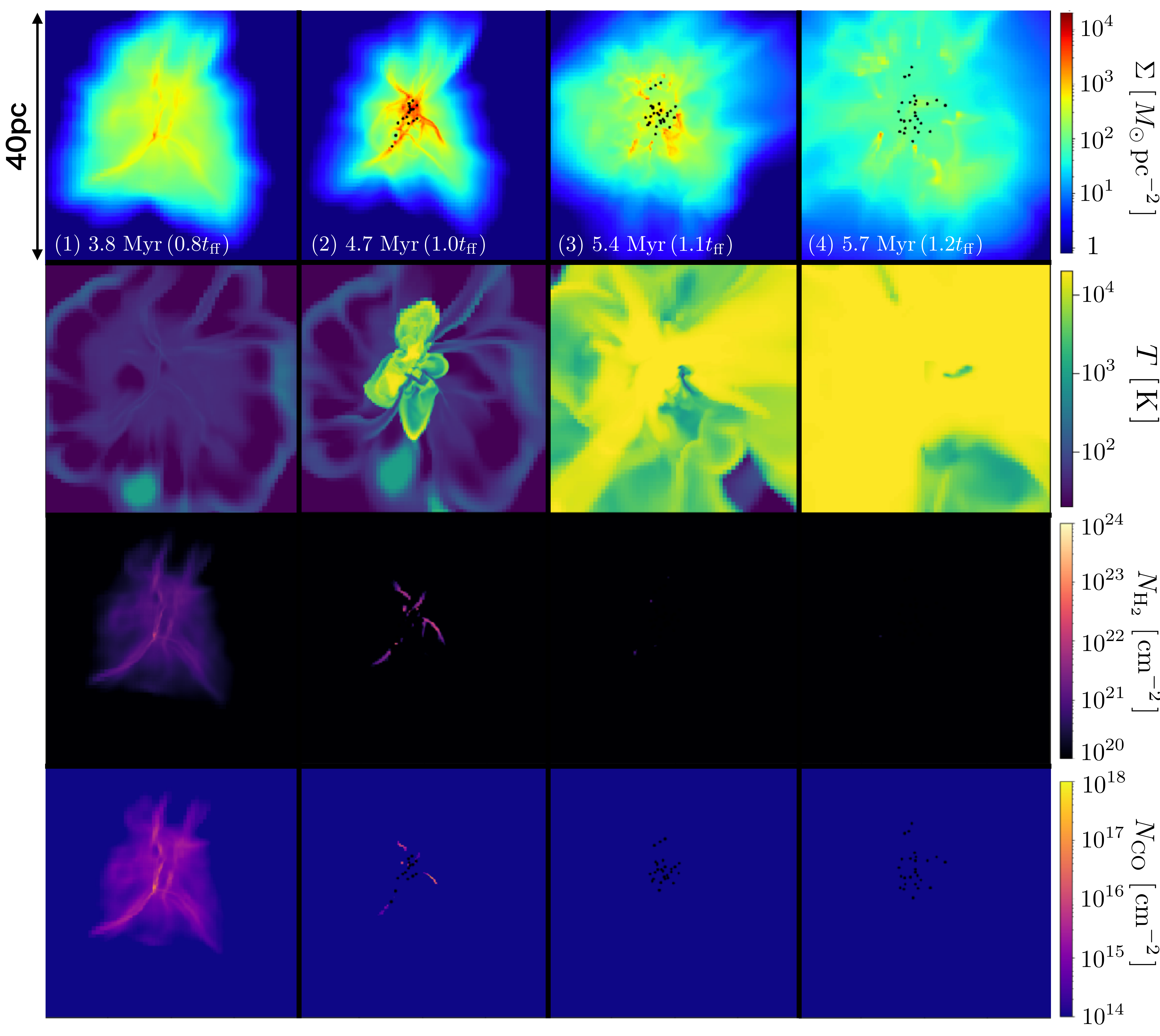}
 \end{center}
 \caption{ 
 Same as Figure \ref{zu1}, but for the cloud with $(M_{\rm cl}, R_{\rm cl}, Z) = (10^5 ~M_{\odot}, 20~{\rm pc}, 10^{-2}~Z_{\odot})$.
 The color scales for the column densities of $\rm H_2$ and CO molecules are different from those in Figure \ref{zu1}.
  }
 \label{zu2}
 \end{figure*}
 \begin{figure*}
 \begin{center}
 \includegraphics[width=130mm]{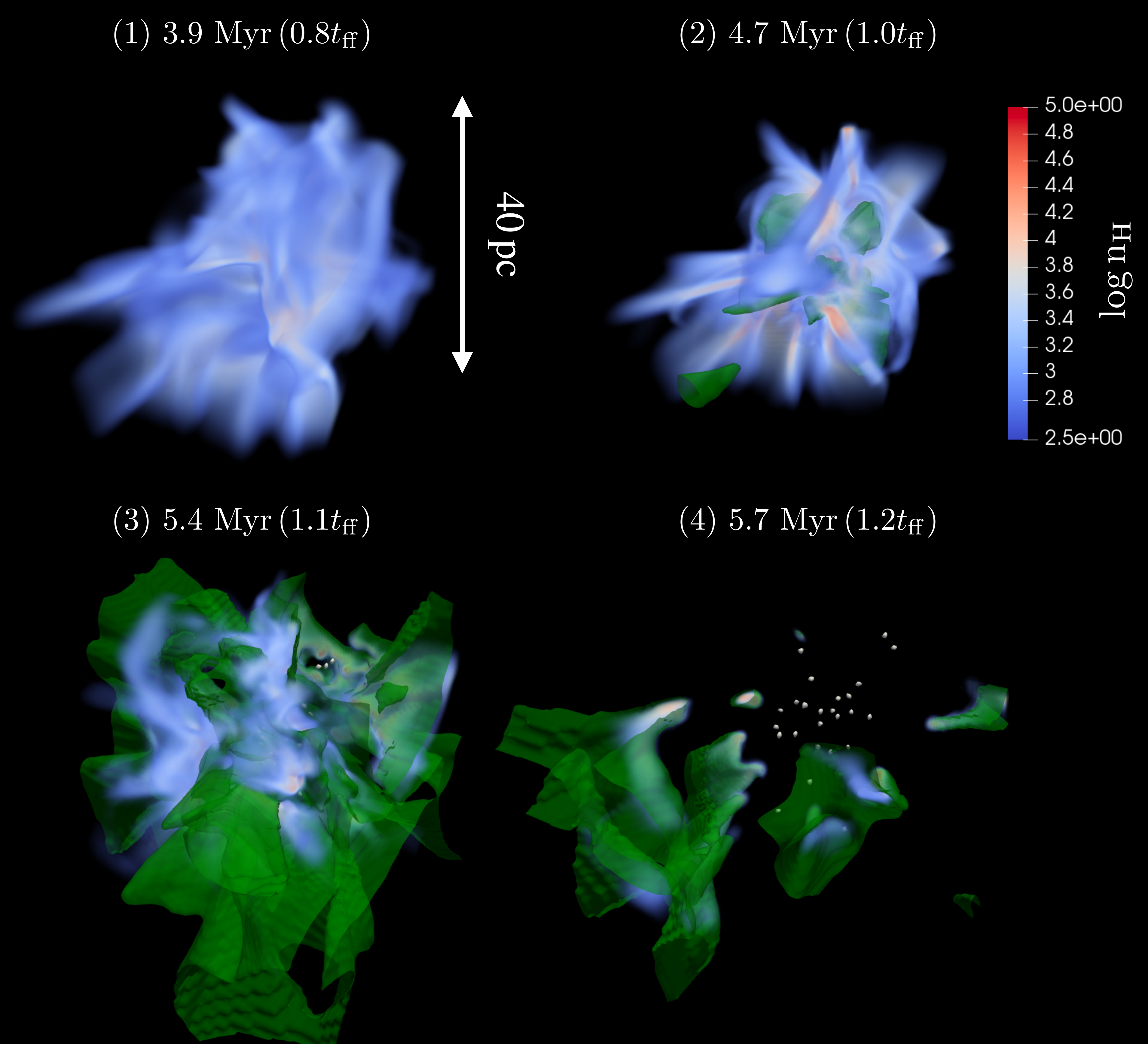}
 \end{center}
 \caption{  
 Same as Figure \ref{zu200108}, but for the cloud with $(M_{\rm cl}, R_{\rm cl}, Z) = (10^5 ~M_{\odot}, 20~{\rm pc}, 10^{-2}~Z_{\odot})$.
 Each panel shows the snapshot at the same epoch as in Figure \ref{zu2}.
   }
 \label{zu200108_2}
 \end{figure*}

Next, we consider the low-metallicity case (M5R20Z-2 in Table \ref{tab1}) with $Z=10^{-2}~Z_{\odot}$ to examine the metallicity dependence of the evolution. Figures \ref{zu2} and \ref{zu200108_2} illustrate the evolution for this case in the same styles as in Figures \ref{zu1} and \ref{zu200108}, respectively. 
In the early epoch, the turbulent motion gives the dominant contribution. The overall evolution looks similar to that in the fiducial case with $Z=1~Z_{\odot}$ until the first star particle forms at $4.0~{\rm Myr}$ (Fig \ref{zu2}-1, \ref{zu200108_2}-1), even though the metal cooling is inefficient.
Once star formation occurs, H{\sc ii} regions start to expand into low-density regions at $4.7~{\rm Myr}$ (Fig \ref{zu2}-2, \ref{zu200108_2}-2).
Unlike the case with $Z=1~Z_{\odot}$, dense filaments are easily photoionized, and they cannot survive around the H{\sc ii} region.
The neutral gas is pushed out all at once, and star formation is completely quenched at $5.4~{\rm Myr}$ (Fig \ref{zu2}-3, \ref{zu200108_2}-3).
Finally, the H{\sc ii} regions spread over the entire cloud and complete the photoevaporation of the gas inside the cloud at $5.7~{\rm Myr}$ (Fig \ref{zu2}-4, \ref{zu200108_2}-4).
 
Only small amounts of $\rm H_2$ and CO molecules form along the filaments in the early phase of the cloud collapse, but the stellar FUV photons easily dissociate them. As shown in Figure \ref{zu2}, $\rm H_2$ and CO molecules completely disappear after one free-fall time unlike for the cloud with $Z=1 ~Z_{\odot}$ (see Fig \ref{zu2}-2). The cloud becomes dark in CO lines soon after the star formation begins. 
  
In the bottom panel of Figure \ref{zu1.2}, we show the time evolution of the mass in each component for the case with $Z=10^{-2}~Z_{\odot}$. We see that the $\rm H_2$ mass is much lower than that for the case with $Z=1~Z_{\odot}$ throughout the evolution.
Since $\rm H_2$ formation on the dust grains is inefficient due to lower metallicity (and hence lower dust abundance), most of the gas remains atomic from the beginning of the simulation. In this case, the star cluster formation occurs in an almost atomic cloud rather than in a molecular cloud \citep{2012ApJ...759....9K}.
Meanwhile, the ionized gas mass increases until the ionized gas is finally evaporated from the cloud. The star formation stops at $5~{\rm Myr}$, with the duration of star formation $\sim 1~{\rm Myr}$, much shorter than in the case with $Z=1~Z_{\odot}$.
The SFE is 5 percent and smaller than that for $Z=1~Z_{\odot}$ by a factor of $\sim 3$.
\citet{2019MNRAS.489.1880H} also performed one simulation run assuming $Z=1/40 ~ Z_{\odot}$, and they obtained a similar reduction rate of the SFE.

\subsubsection{Radiation pressure on dust grains}\label{radiative_feedback_on_GMCs}
 \begin{figure}
 \begin{center}
 \includegraphics[width=\columnwidth]{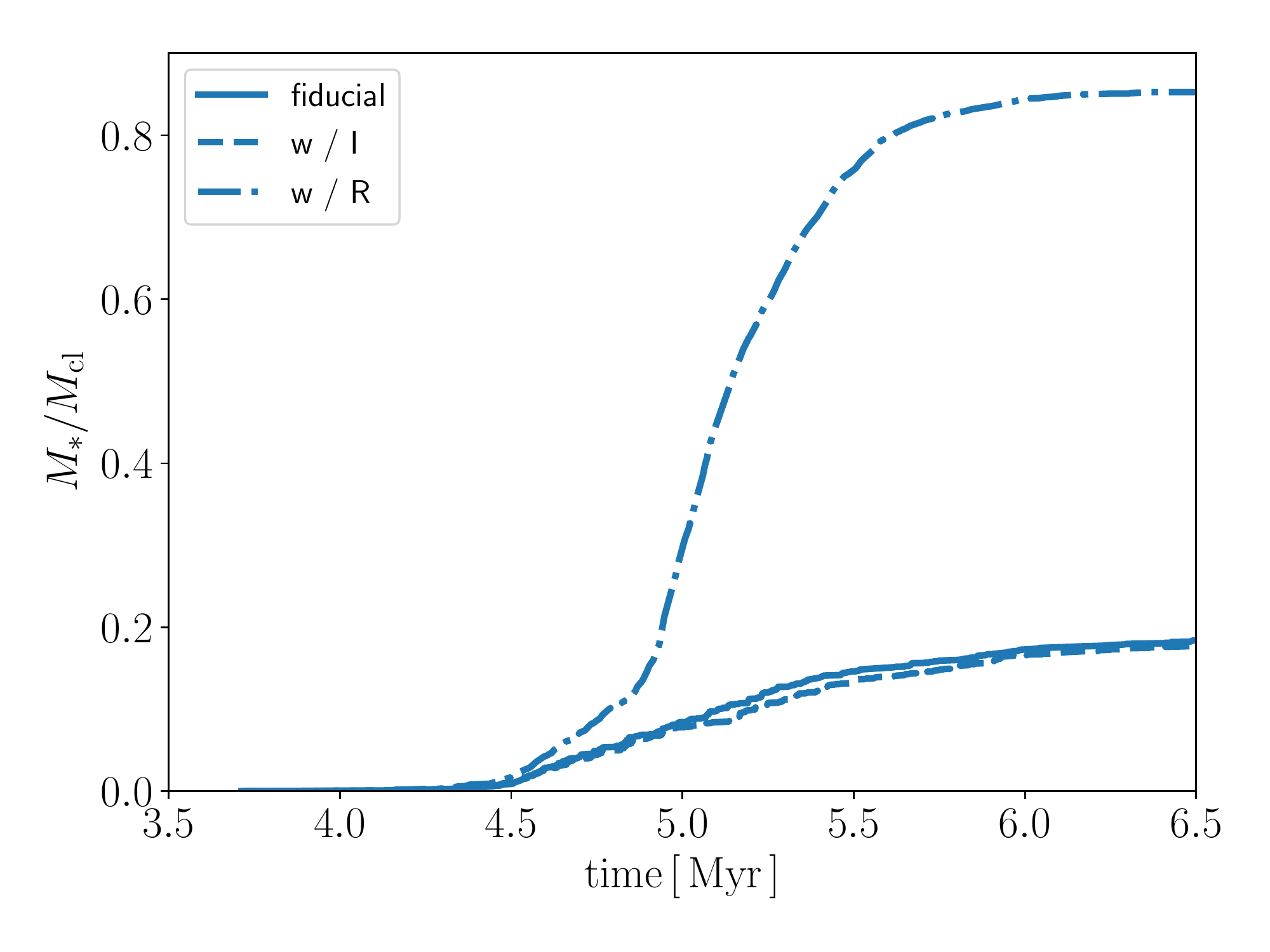}
 \end{center}
 \caption{ 
 Impact of photoionization and radiation pressure on star formation.
 Each line shows the time evolution of the total stellar mass in the cloud with $(M_{\rm cl}, R_{\rm cl}, Z) = (10^5 ~M_{\odot}, 20~{\rm pc}, 1~Z_{\odot})$.
 The solid line represents the fiducial case (M5R20Z0) as shown in Figure \ref{zu1.2}.
 Other lines shows the cases where we only consider either photoionization (dashed line, M5R20Z0PI) or radiation pressure (dot-dashed line, M5R20Z0RP).
 }
 \label{mass_feedback}
 \end{figure}

In Section \ref{1Zsun} and \ref{low_metallicity_cloud}, we have seen that the expansion of H{\sc ii} regions quenches star formation while the radiation pressure on dust grains also affects the gas motion. 
Here, we investigate the individual effects of photoionization and radiation pressure on star formation separately.
Figure \ref{mass_feedback} shows the time evolution of total stellar masses in the cases where we only consider either photoionization (M5R20Z0PI, dashed line) or radiation pressure (M5R20Z0RP, dot-dashed line).
With only photoionization, star formation starts at $\sim 4.5~{\rm Myr}$, and continues until $\sim 7~{\rm Myr}$.
The star formation history is almost the same as in the fiducial case, where we include both photoionization and radiation pressure effects. 
On the other hand, with only radiation pressure effect the SFE is very high exceeding 0.8. 
This result is consistent with the simulation of \citet{2016ApJ...829..130R} at $\alpha = 0.4$. This demonstrates that the photoionization feedback plays the leading role in regulating star formation.

Next, we discuss why the radiation pressure plays only a secondary role in our simulations.
As given in Equation \eqref{1.1}, the kinetic energy of turbulent motion is less than the gravitational binding energy in the initial condition.
Then, the star clusters appear in the central region of the cloud, and the gas accretes onto the central star-forming regions in a spherical symmetric fashion, as shown in Figure \ref{zu1}.
In such a situation, the supply of gas is suppressed if the momentum flux of radiation exceeds that of the inflow \citep{1987ApJ...319..850W},
\begin{eqnarray}
	\frac{L_*}{4 \pi R^2 c} > \rho u^2, \label{momentum_flux}
\label{condition_acc_org}
\end{eqnarray} 
where $L_*$, $R$ and $u$ are total stellar luminosity, the distance from the star-forming region and the infall velocity of the gas.
Here, we assume that the infall velocity to the star cluster is the free-fall velocity as $u = \sqrt{2 G M_*/R}$ where $M_*$ is the total stellar mass.
Using the relation between the stellar mass and luminosity $L_* = l_* M_*$ where the factor of proportionality $l_* = 1.2\times10^3 ~L_{\odot} M_{\odot}^{-1}$ estimated in Section \ref{radsource}, Equation \eqref{momentum_flux} becomes
\begin{eqnarray}
 R < \frac{l_*}{8 \pi G c \mu m_{\rm H} n_{\rm H}} = 0.65 ~{\rm pc} \left( \frac{n_{\rm H}}{10^{4}~{\rm cm^{-3}}} \right)^{-1}, 
\label{condition_acc}
\end{eqnarray}
where $\mu$ is the mean molecular weight. 
Equation \eqref{condition_acc} shows that radiation pressure is effective only to very dense filaments $(n_{\rm H} \sim 10^{4}~{\rm cm^{-3}})$ falling within $\sim 1~{\rm pc}$ of a star cluster. Thus the impact of radiation pressure depends on the distribution of the gas and stars. In our simulations, the dense filaments fragment into clumps, which then collapse into stars before reaching $\sim 1~{\rm pc}$ from pre-existing massive stars. This explains why the radiation pressure effect is so weak. Note that the above argument is only applicable for limited cases where the whole cloud collapses nearly spherically owing to relatively weak initial turbulent support. \citet{2018ApJ...859...68K} showed that with only the radiation pressure feedback the SFE is reduced to $\simeq 0.2$ assuming $\alpha_0 = 2.0$ instead of our value $\alpha_0 = 0.5$. In such a case, the whole cloud is more violently disrupted by the stronger turbulence and the evolution is no longer approximated with the spherical collapse as in Equation \eqref{condition_acc_org}.

If $Z \ll  1~Z_{\odot}$, the cloud is not optically thick (see also, sec. \ref{optical_depth_of_shielding}), and the radiation pressure has little impact. Thus, we take only the photoionization feedback into account in developing a semi-analytical model in Section \ref{semi_analytical_model}.

\subsection{Variations among models with different cloud properties}
\label{propGMC}

\subsubsection{Star formation efficiency}\label{SFE_GMC}

 \begin{figure}
 \begin{center}
 \includegraphics[width=\columnwidth]{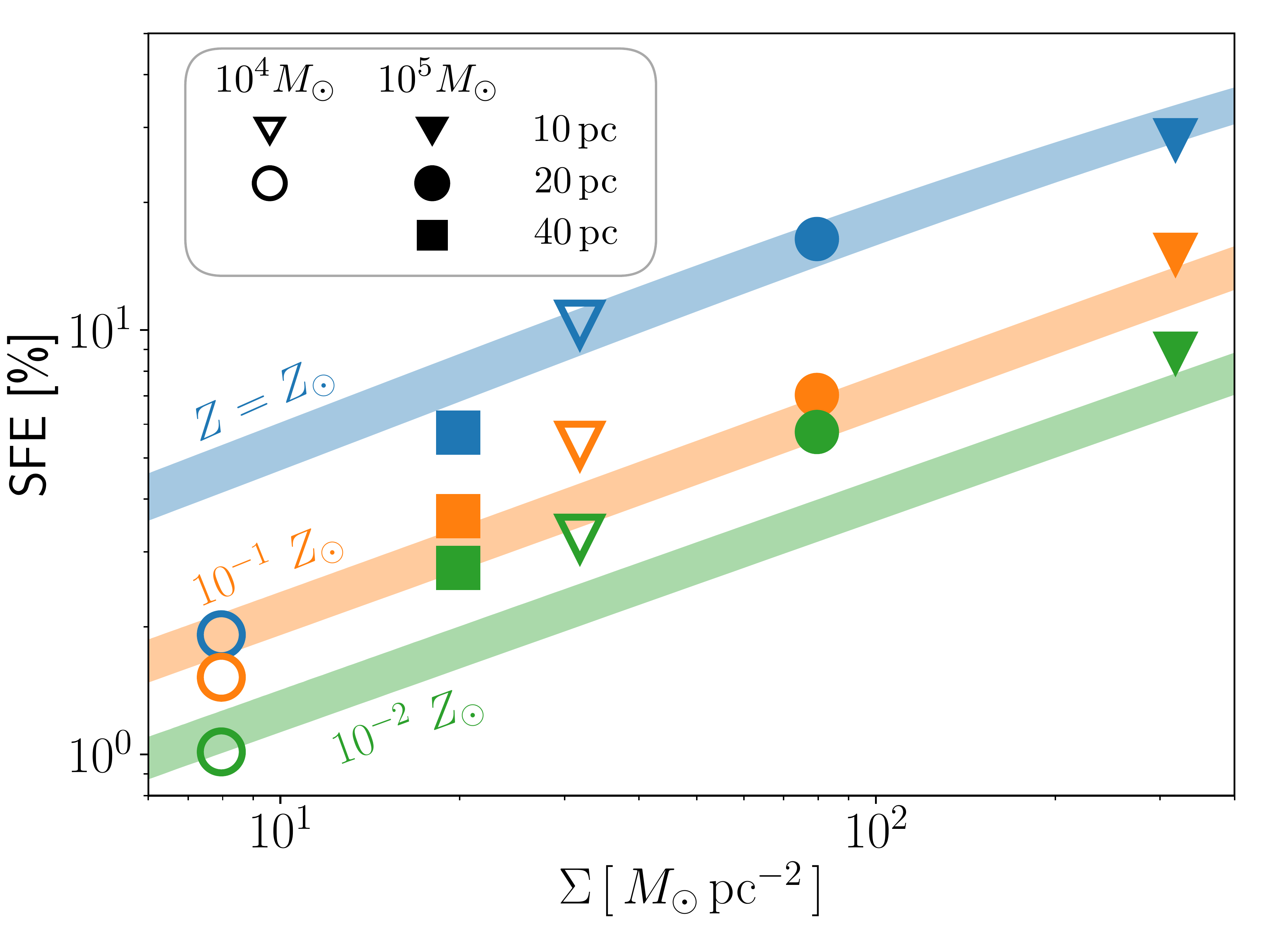}
 \end{center}
 \caption{The star formation efficiency (SFE) as a function of the cloud surface density $\Sigma$, mass $M_{\rm cl}$,  and metallicity $Z$. The symbols represent the simulation results for the clouds with the different radii $R_{\rm cl} = 10~{\rm pc}$ (square), $20~{\rm pc}$ (circle), and $40~{\rm pc}$ (triangle), and with the different masses $M_{\rm cl} = 10^{5}~M_{\odot}$ (filled) and $10^{4}~M_{\odot}$ (open). The shaded stripes show our analytical estimate of the SFE for the clouds with mass between $M_{\rm cl} = 10^{4}$ and $10^{5}~M_{\odot}$ given by Equation \eqref{1228.9}. The different colors represent different metallicities of $Z=1~Z_{\odot}$ (blue), $10^{-1}~Z_{\odot}$ (orange) and $10^{-2}~Z_{\odot}$ (green) for both the symbols and stripes. 
 }
 \label{zu3}
 \end{figure}

For the models with various $M_{\rm cl}$, $R_{\rm cl}$ and $Z$, we obtain the SFE $\epsilon_* \equiv M_{*}/M_{\rm cl}$ from the simulations, where $M_*$ is the final total stellar mass.
Figure \ref{zu3} shows the SFE as a function of the initial surface density of the clouds.
The SFE increases with the surface density, consistently with previous studies \citep[e.g.,][]{2016ApJ...829..130R, 2018ApJ...859...68K, 2019MNRAS.489.1880H}.
For the case with $Z=1~Z_{\odot}$, the SFE increases from 2 to 30 percent as $\Sigma$ increases from $8$ to $300~{M_{\odot} {\rm pc^{-2}}}$.
At each metallicity, the SFE is well approximated by a power-law function of the surface density except for the case with the lowest surface density ($\Sigma \sim 8~{M_{\odot} {\rm pc^{-2}}} $). 
At $\Sigma \sim 8~{M_{\odot} {\rm pc^{-2}}} $, a single or a few star clusters easily photoionize the entire region of the cloud. In these cases, the SFEs do not follow the power-law relations, regardless of metallicity.

Note that the EUV/FUV emissivity of an individual cluster particle should also fluctuate when the mass is small, owing to the stochastic sampling of high-mass stars. \citet{2016ApJ...819..137K} showed that the assumption of a constant emissivity was valid for $M_* \gtrsim 10^{4}~M_{\odot}$, but the stochasticity appears if the cluster mass was lower. In low-mass clusters with $M_{* } \lesssim 10^2~M_{\odot}$, furthermore, the averaged emissivity of EUV/FUV photons should be smaller than the constant value we currently assume because of the scarcity of high-mass stars \citep{Inoguchi20}. We expect that, for the low-surface density clouds, the above effects enhance the SFEs and reduce the deviations from the power-law relation on average, albeit with some scatters.

The SFE also increases with the metallicity.
This is because the photoionization feedback that disrupts the star-forming clouds
becomes stronger at lower metallicity from the following two reasons.
Firstly, the temperature in the H{\sc ii} regions, 
where the main coolant is the metal lines of O{\sc ii} and O{\sc iii} \citep{2006agna.book.....O},
is higher in lower-metallicity cases: its typical value increases from $9 \times 10^3~{\rm K}$ at $Z=1 ~Z_{\odot}$ to $2.0 \times 10^{4}~{\rm K}$ at $Z=10^{-2}~Z_{\odot}$.
The resultant higher pressure causes more rapid expansion of the H{\sc ii} regions and thus stronger feedback to the star-forming clouds.
Secondly, the ionizing photons are less attenuated by dust grains in lower-metallicity cases.
As discussed in Section \ref{fiducial_case}, in the case with $Z=1~Z_{\odot}$, the star-forming filaments can survive even after the H{\sc ii} regions begin to expand because dust grains shield the neutral gas from the photoionization.
For the case with $Z=10^{-2}~Z_{\odot}$, however, the filaments are easily photoionized and soon disrupted. 
From the reasons above, the duration of star formation is short, and thus the SFE is low in lower-metallicity environments. 
The dust shielding of the filaments will be discussed analytically in Section \ref{optical_depth_of_shielding}.

\subsubsection{Lifetime of the clouds}
 \begin{figure}
 \begin{center}
 \includegraphics[width=\columnwidth]{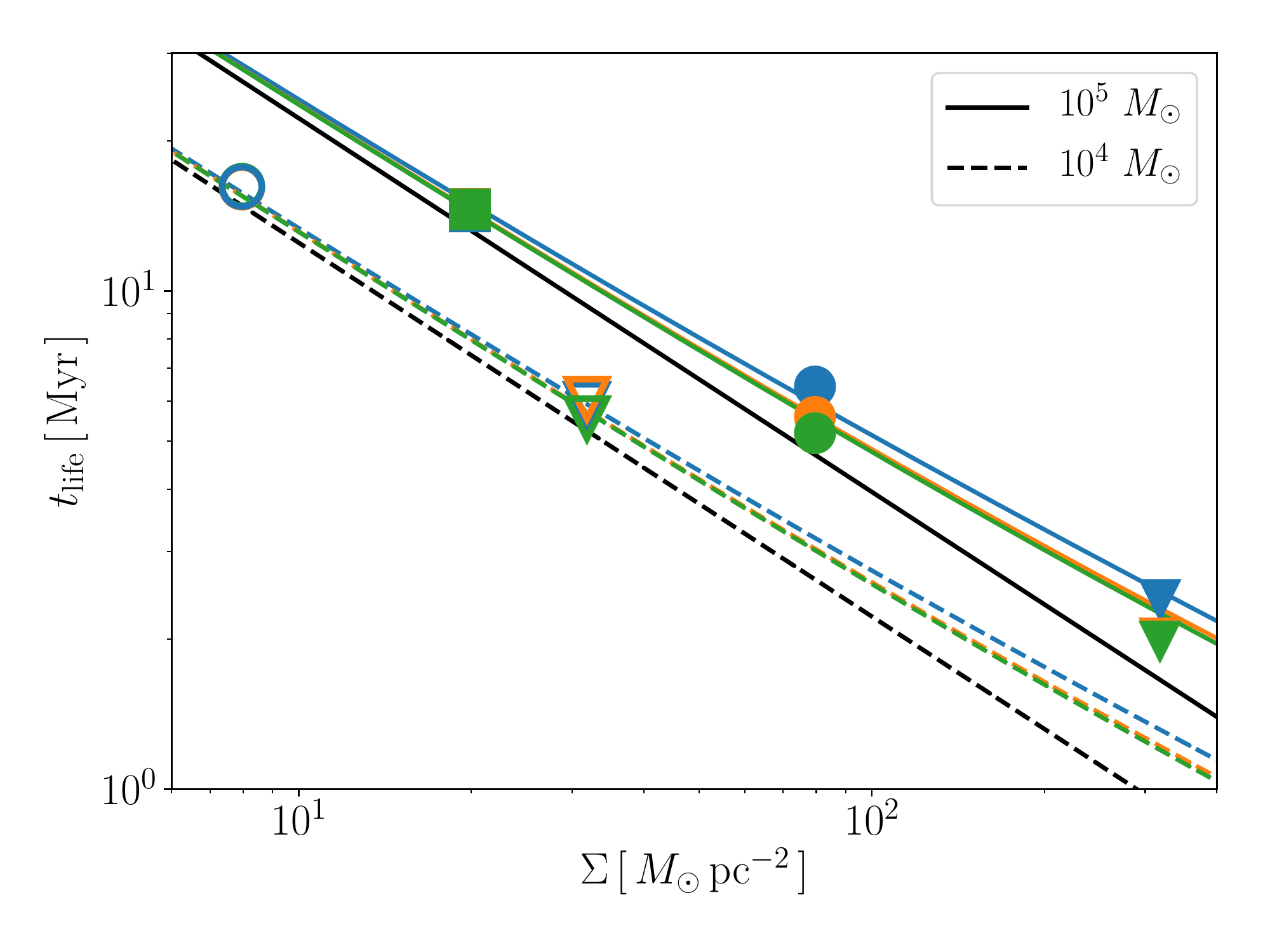}
 \end{center}
 \caption{
 The lifetime of star-forming clouds as a function of the surface density.
We use the same symbols and colors as in Figure \ref{zu3} to indicate the different conditions of $R_{\rm cl}$, $M_{\rm cl}$ and $Z$.
The analytical estimations of the lifetime for the clouds with $M_{\rm cl} = 10^{5}~M_{\odot}$ (solid) and $10^{4}~M_{\odot}$ (dashed) 
and $Z=1~Z_{\odot}$ (blue), $10^{-1}~Z_{\odot}$ (orange) and $10^{-2}~Z_{\odot}$ (green) are also shown with the lines, 
which are calculated from $t_{\rm life} = t_{\rm ff} + t_{\rm HII}$ with the free-fall time of the clouds $t_{\rm ff}$ (shown with the black lines)
and the analytical estimate of the duration of star formation $t_{\rm HII, cl}$ (Eq. \ref{0109}).
 }
 \label{zu335}
 \end{figure}

In Figure \ref{zu335}, we show the lifetime of cluster-forming clouds $t_{\rm life}$ as a function of the surface density. In the numerical simulations, we define the cloud lifetime as elapsed time since the start of the simulations until the total stellar mass reaches 90 percent of the final value. 
In low surface-density cases with $\Sigma \lesssim 30~{M_{\odot} {\rm pc^{-2}}}$,
the duration of the star formation is much shorter than the free-fall time of the clouds, and hence the lifetime is almost equal to the latter.
In high surface-density cases with $\Sigma \gtrsim 70~{M_{\odot} {\rm pc^{-2}}}$, however,
the lifetime is slightly longer than the free-fall time because of the longer expansion time of the H{\sc ii} regions.
In some cases, the lifetimes of clouds can be longer than that of massive stars ($\sim 3 ~ {\rm Myr}$).
However, the duration time from the onset of star formation to the disruption is much shorter than the lifetime of massive stars in all the cases. 
Therefore our modeling of the luminosity from a sink particle can be justified.

\subsection{Evolution of cluster particles}
\label{properties_of_star_clusters}

 \begin{figure}
 \begin{center}
 \includegraphics[width=\columnwidth]{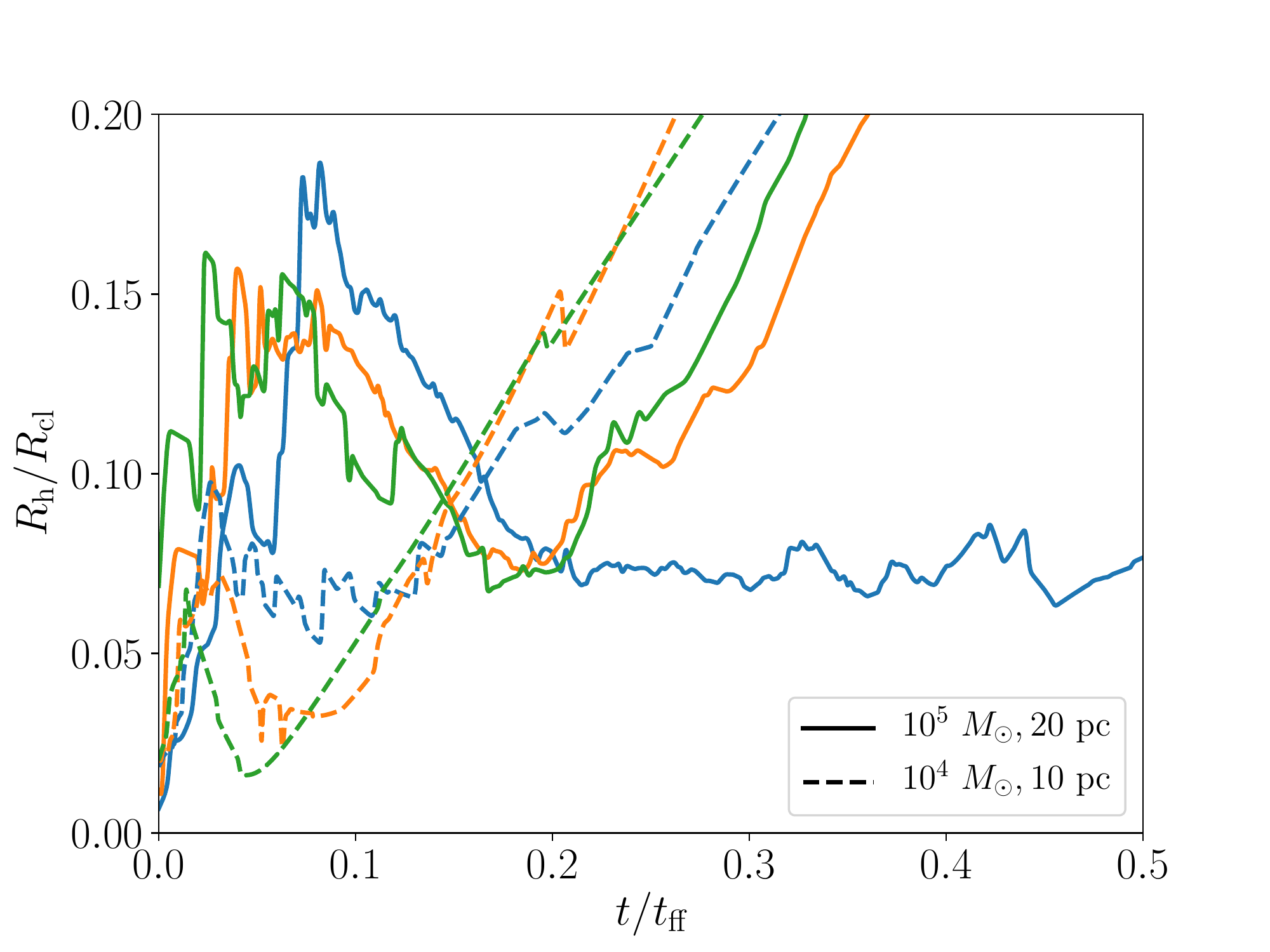}
 \end{center}
 \caption{The time evolution of the half-mass radius $R_{\rm h}$ of the cluster particles for the cases with  $(M_{\rm cl}, R_{\rm cl}) = (10^{5}~{M_{\odot}}, 20~{\rm pc})$ ("M5R20" models, solid line) and $(10^{4}~{M_{\odot}}, 10~{\rm pc})$ ("M4R10" models, dashed line) as a function of the time normalized by the free fall time of the clouds $t_{\rm ff}$. The initial surface density $\Sigma$ is $80~M_\odot~{\rm pc}^{-2}$ for the former cases and $32~M_\odot~{\rm pc}^{-2}$ for the latter. The line colors indicate the metallicity: $Z=1~Z_{\odot}$ (blue), $10^{-1}~Z_{\odot}$ (orange) and $10^{-2}~Z_{\odot}$ (green).}
 \label{zu_sink}
 \end{figure}
 \begin{figure}
 \begin{center}
 \includegraphics[width=\columnwidth]{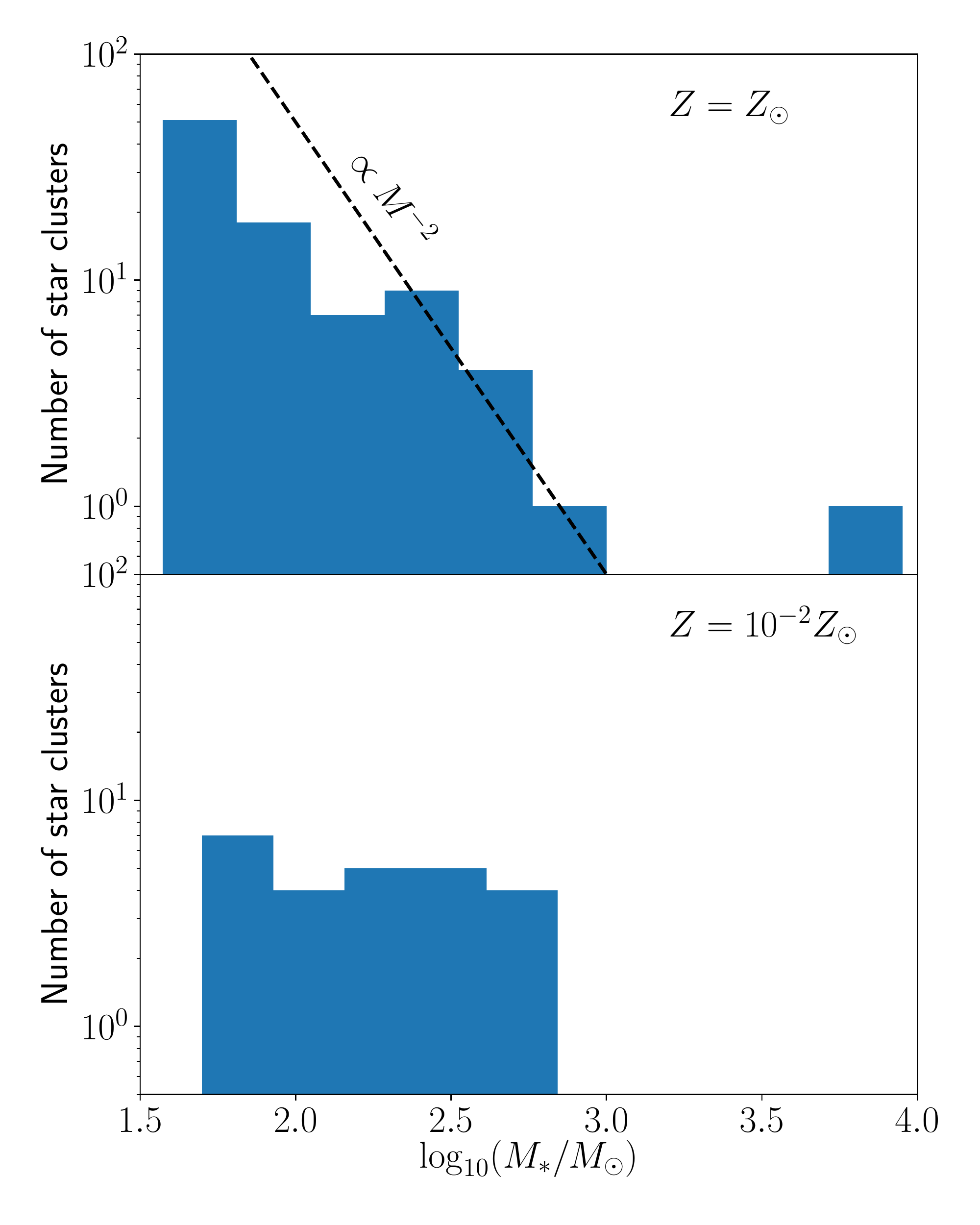}
 \end{center}
 \caption{
 The mass distribution of the cluster particles at the end of the simulations.
 Top and bottom panels show the cases with $(M_{\rm cl}, R_{\rm cl}, Z) = (10^{5}M_{\odot}, 20{\rm pc}, 1 ~Z_{\odot})$ and $(10^{5}M_{\odot}, 20{\rm pc}, 10^{-2}~Z_{\odot})$.
The dashed line indicates the relation $\propto M_*^{-2}$.
 }
 \label{zu_mass_dis}
 \end{figure}

The early evolution of newborn star clusters depends on the initial cloud surface-density and metallicity.
In a more compact and higher-metallicity cloud, the SFE and the total stellar mass are higher. Thus the star cluster formed there tends to remain gravitationally bound.
In our case, the star clusters remain gravitationally bound until the end of the simulation only in cases with metallicity $Z=1~Z_{\odot}$ and with initial surface density $\Sigma \gtrsim 80~{M_{\odot}{\rm pc^{-2}}}$.  
In Figure \ref{zu_sink}, the time evolution of the half-mass radius of the star clusters formed in our simulations is shown in some cases.
In the case with $(M_{\rm cl}, R_{\rm cl}, Z) = (10^{5}~{M_{\odot}}, 20~{\rm pc}, 1 Z_{\odot}) $, where $\Sigma \simeq 80~{M_{\odot}{\rm pc^{-2}}}$, the half-mass radius remains constant at $1.5~{\rm pc} ~(\sim 0.075~R_{\rm cl})$ after the quenching of the star formation, and the star clusters remain gravitationally bound. 
We see that even with the same $\Sigma$, the clusters finally dissolve at lower metallicities.
In other cases with $\Sigma \simeq 32~{M_{\odot}{\rm pc^{-2}}}$, the half-mass radii increase monotonously and the star clusters become unbound regardless of $\Sigma$. In these cases, the timescale of dissolving the clusters corresponds to that of the H{\sc ii}-region expansion.
Before the disruption of the clouds, the star cluster is trapped in the central region due to the gravitational potential of the gas.
However, it becomes shallower after the H{\sc ii}-region expansion and the star cluster comes apart if the gravitational binding is not sufficient, as shown in Figure \ref{zu_sink}. 
The above trend may be consistent with the high "infant mortality" of embedded clusters \citep[e.g.,][]{Lada03}.
Note that, individual stars modeled in a sink particle can also be dispersed as a result of the cloud photoevaporation and the properties of star clusters change via the interaction between individual stars \citep{2001MNRAS.321..699K, 2013ApJ...764...29B, 2017A&A...597A..28B}. Following these processes is out of the scope in this work.

In Figure \ref{zu_mass_dis}, the mass distribution of the cluster particles are shown for cases with 
metallicities $Z=1~Z_{\odot}$ and $10^{-2}~Z_{\odot}$. 
We see that the mass spectrum at $Z=10^{-2}~Z_{\odot}$ is much flatter than that at $Z=1~Z_{\odot}$. The mass distribution at $Z=1~Z_{\odot}$ is well fitted by a power-law $dN/dM_*\propto M_*^{-2}$, which is consistent with observed mass functions of nearby open and young massive clusters \citep[e.g.,][]{2010ARA&A..48..431P, 2019ARA&A..57..227K}. 
We argue that this dependence on metallicity comes from how tight the system is gravitationally bound. In the case with $Z=10^{-2}~Z_{\odot}$, the cluster particles are only loosely bound at birth and immediately dissolve through the cloud dispersal as described above. Mergers between cluster particles hardly occur. We have confirmed that, although not presented, the flat mass distribution is broadly observed in the other cases where the particles become unbound after the cloud dispersal. At $Z=1~Z_{\odot}$, on the other hand, the cluster particles are more tightly bound for long enough time, resulting in more frequent merger events. It appears that the power-law mass distribution is a result of such difference in the evolution.

\section{Semi-analytical argument}\label{semi_analytical_model}

\subsection{Dependency of SFE on physical conditions of clouds}\label{Rn_in_starforming_region}

As shown in the previous section, the SFE and the cloud lifetime vary with the initial surface density and metallicity in our simulations. In previous works, semi-analytical models have been developed to interpret the simulation results, particularly the variation of the SFE.
\citet{2018ApJ...859...68K} considered the total momentum of the neutral and ionized gases ejected from photoevaporating clouds, using their simulation results. \citet{2019MNRAS.489.1880H} evaluated the SFE using the star formation rate and the sound crossing time of H{\sc ii} regions calibrated with their simulation results, although limited only to the case of $Z=1~Z_{\odot}$.
In this section, we construct a new semi-analytical model in a different approach, paying special attention to the metallicity dependence of photoionization feedback. We consider the propagation time of ionizing front $t_{\rm HII}$ as the duration of star formation because the star formation continues until the expanding H{\sc ii} bubbles terminate gas supply by inflows to dense star-forming regions. Then, the total stellar mass $M_* = \epsilon_* M_{\rm cl}$ is obtained by multiplying $t_{\rm HII}$ with the star formation rate $\dot M_*$.
In our model, $t_{\rm HII}$ also depends on the total stellar mass of radiation sources.
Besides, the star formation rate $\dot M_*$ varies with metallicity because the cooling time of gas becomes longer in low-metallicity environments.
Therefore, we need to calculate the SFEs $\epsilon_{*}$, the propagation time $t_{\rm HII}$, and the star formation rate $\dot M_*$ consistently.
In the following, we consider the conversion rate of the neutral gas to stars in order to formulate the star formation rate $\dot M_*$ in Section \ref{fracofstarforming}.
We then construct a model for the SFE $\epsilon_{*}$ and the propagation time $t_{\rm HII}$ in Section \ref{SFEs_model}.

\subsubsection{Fraction of star-forming gas around star clusters}
\label{fracofstarforming}

 \begin{figure}
 \begin{center}
 \includegraphics[width=\columnwidth]{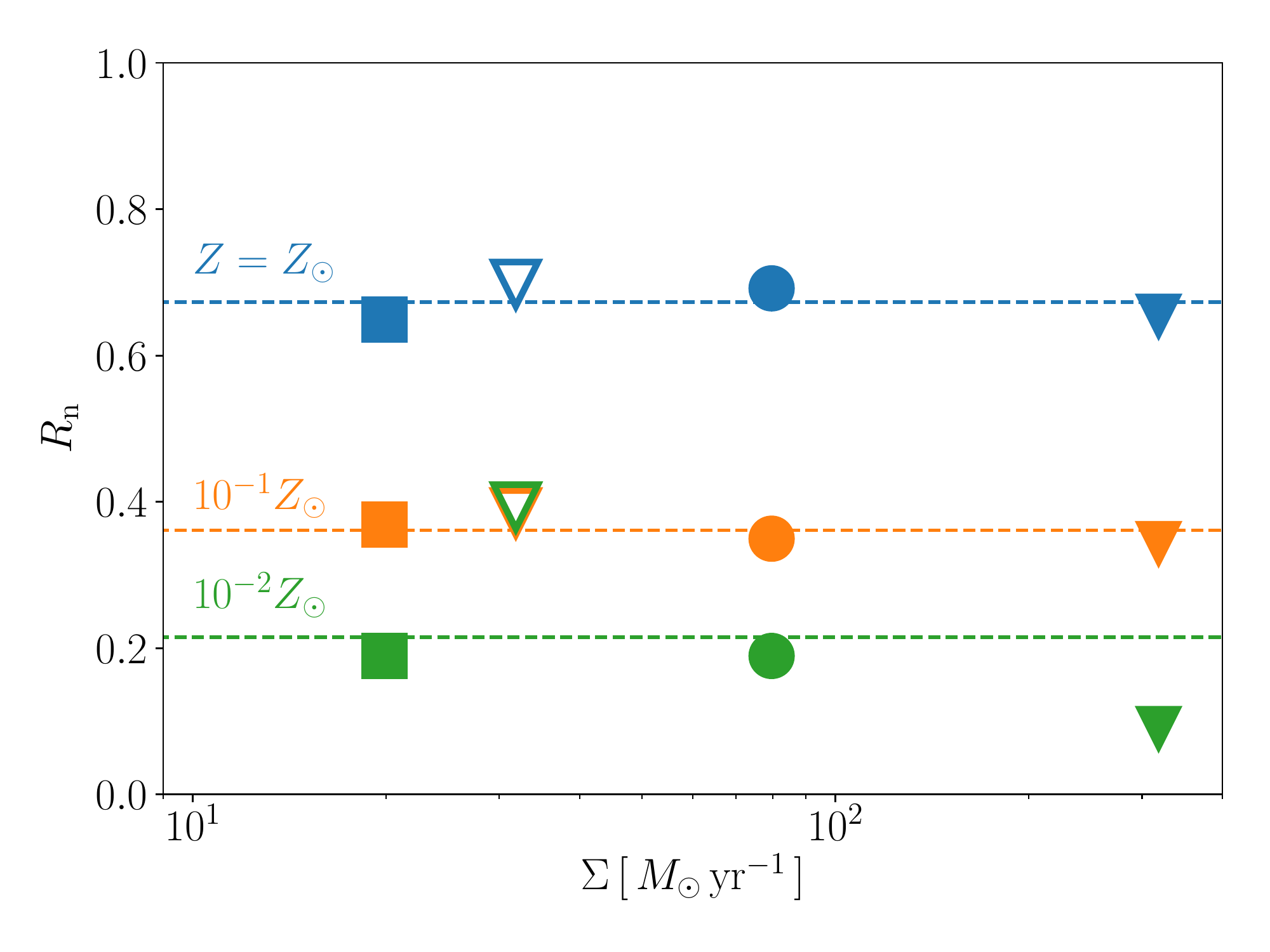}
 \end{center}
 \caption{
 The average fraction of the mass of the cold gas ($T < 200~{\rm K}$) $R_{\rm n}$ as given in Equation \eqref{EqRn}.
 The symbols and colors are same as Figure \ref{zu3}.
The dashed lines shows the averaged value of $R_{\rm n}$ with $\Sigma > 10~{M_{\odot} {\rm pc^{-2}}}$ (see text) at each metallicity.
 }
 \label{zu0215_2}
 \end{figure}
 \begin{figure}
 \begin{center}
 \includegraphics[width=\columnwidth]{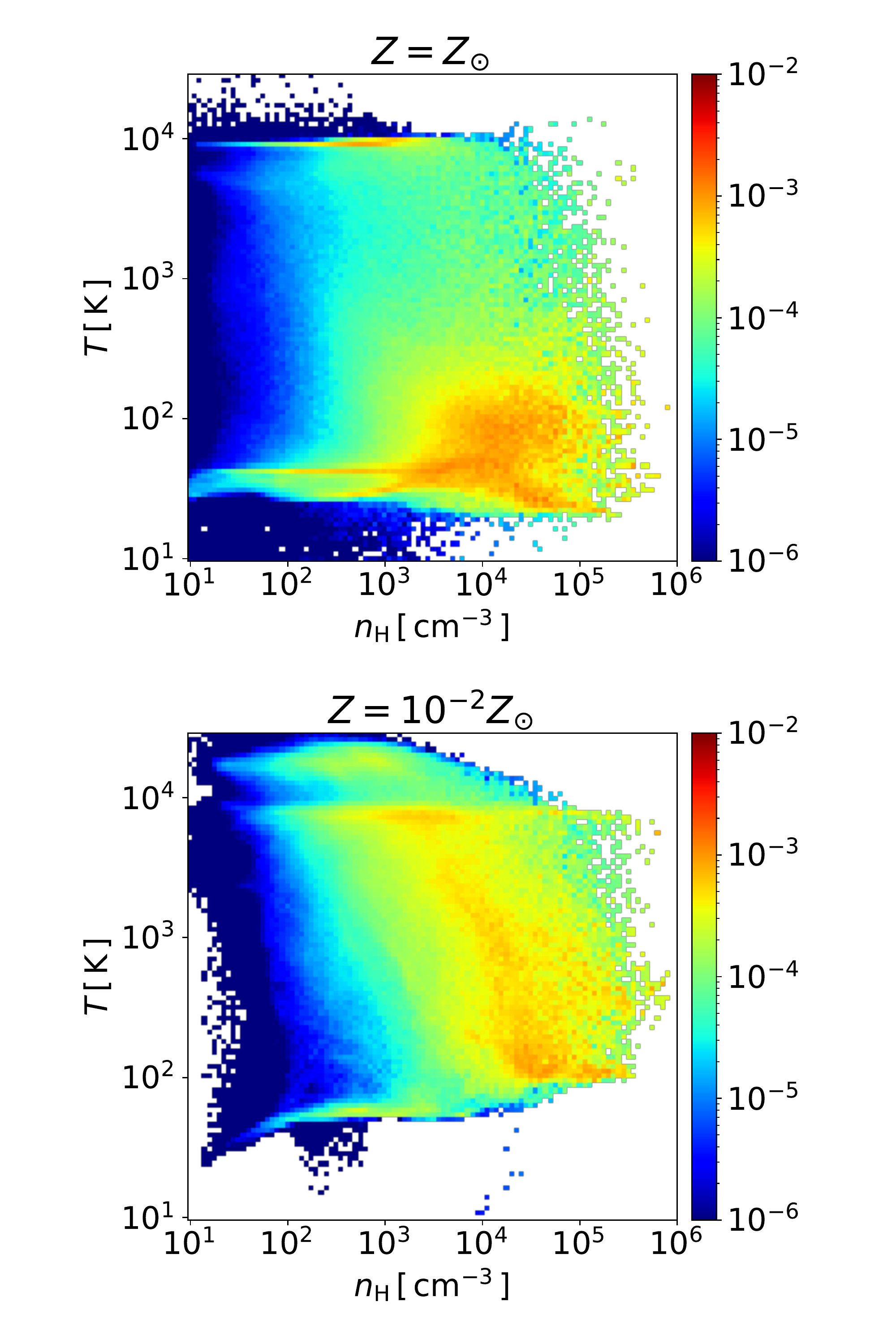}
 \end{center}
 \caption{The gas mass distributions on the $n_{\rm H}- T$ plane within the star-forming region $r_{\rm sf}$ (see text) at the epoch of $t=t_{\rm ff}$.
The top and bottom panels represent the cases with $(M_{\rm cl}, R_{\rm cl}, Z) = (10^{5}M_{\odot}, 20{\rm pc}, 1Z_{\odot})$ and $(10^{5}M_{\odot}, 20{\rm pc}, 10^{-2}Z_{\odot})$.
In each panel the colors show the mass ratio of each bin to the total mass.}
 \label{zu_rhoT}
 \end{figure}

 \begin{figure}
 \begin{center}
 \includegraphics[width=\columnwidth]{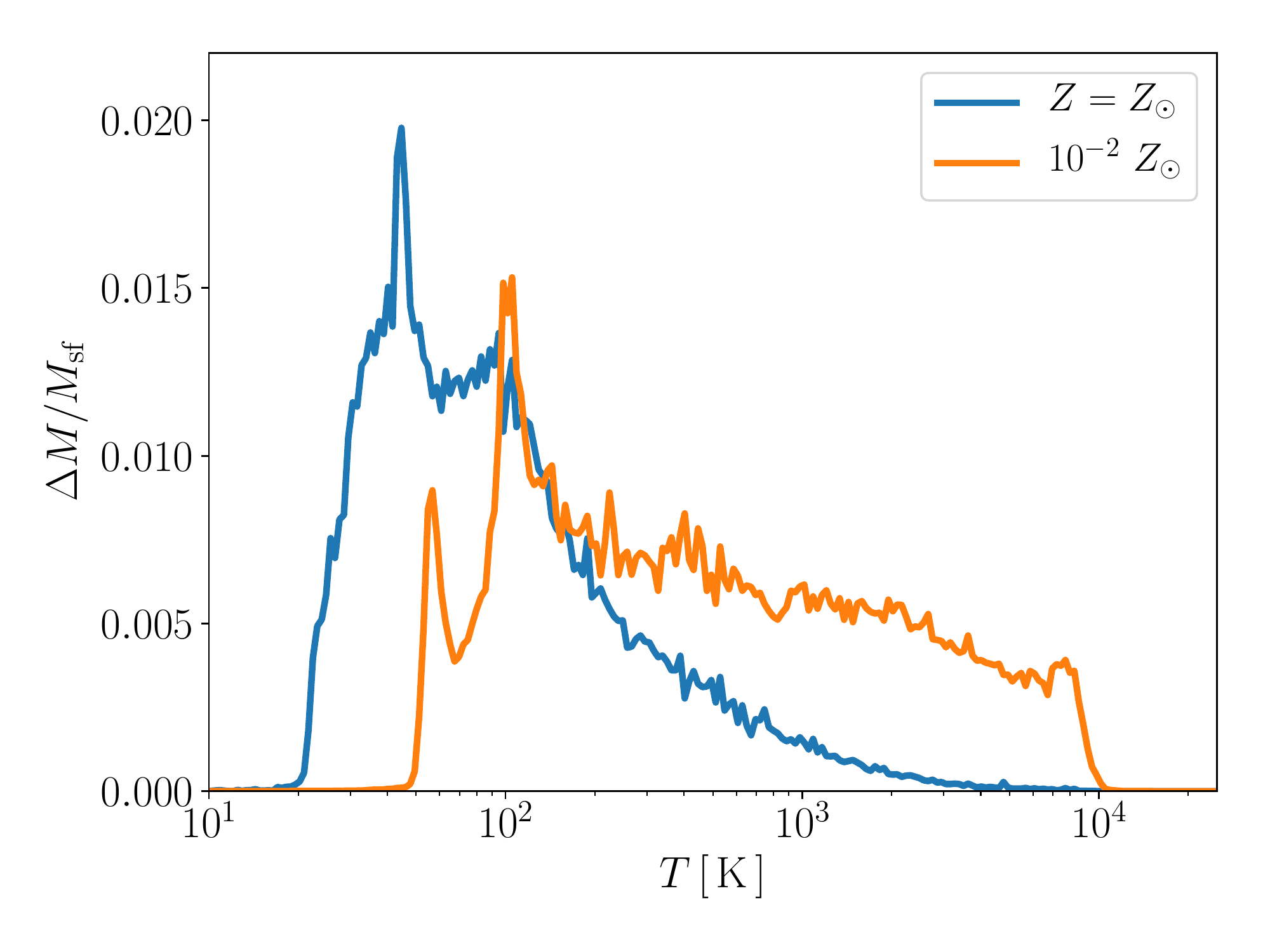}
 \end{center}
 \caption{
The mass distribution against the temperature within the star-forming region $r < r_{\rm sf}$ at the epoch of $t=t_{\rm ff}$.  
Plotted is the mass fraction at each temperature bin $\Delta M / M_{\rm sf}$, where $M_{\rm sf}$ is the total mass contained within the radius $r_{\rm sf}$.
Only the non-ionized gas is considered for the plots.
The blue and orange lines represent the cases with $(M_{\rm cl}, R_{\rm cl}, Z) = (10^{5}M_{\odot}, 20{\rm pc}, 1Z_{\odot})$ and $(10^{5}M_{\odot}, 20{\rm pc}, 10^{-2}Z_{\odot})$.
 }
 \label{zu45.1}
 \end{figure}

We find that only a part of neutral gas around stars is used for the subsequent star formation in our simulations. We thus first estimate the amount of cold gas as a reservoir for the star formation. As shown in Figures \ref{zu200108} and \ref{zu200108_2}, the neutral and ionized regions coexist around star clusters. The neutral gas is vulnerable to compressional heating by the expanding H{\sc ii} bubbles and the feedback by FUV photons is secondary. 
At $Z=1~Z_{\odot}$, those dense regions are shielded from EUV/FUV photons by dust grains (see Sec \ref{optical_depth_of_shielding}), and the photodissociation of molecules and photoelectric heating are inefficient.
At low-metallicities, on the other hand, UV radiation can penetrate the clouds, but the clouds consist of atomic gas originally, and the dust content is low. 
Thus, both the photodissociation and photoelectric heating are not important in the neutral gas around the H{\sc ii} regions.
Main cooling processes are the dust cooling and line emission of CO, C{\sc ii} and O{\sc i}.
From the thermal balance among these processes, the gas temperature can be higher than a few $100 ~{\rm K}$, depending on metallicity. 
With higher temperatures, the collapse of a dense part is delayed until its mass exceeds the higher value of Jeans mass by accreting more surrounding gas. In this case, a warm gas can be evacuated due to photo-ionization instead of collapse. Only the gas with temperature below the threshold value collapses promptly, resulting in star formation. 
Here, we assume the critical temperature $T_{\rm cr}=200~{\rm K}$ above which the gas is supposed to be heated up due to the feedback as described above.

Next, we estimate the amount of star-forming gas.
We consider the region inside the active star-forming radius ($r_{\rm sf}$) defined as the distance between the center of mass and the most distant sink particle.
Then we estimate the typical mass fraction of the cold gas during the star formation by calculating the time average weighted with the star formation rate, 
\begin{eqnarray}
	R_{\rm n} = \frac{\int_{t_{10}}^{t_{90}} \dot M_{*} (M_{\rm n} / M_{\rm gas}) dt }{\int_{t_{10}}^{t_{90}} \dot M_{*} dt}, \label{EqRn}
\end{eqnarray}
where $M_{\rm n}$ and $M_{\rm gas}$ are the cold, i.e., $T < 200~\rm K$, and total gas masses. 
In Equation \eqref{EqRn}, $t_{10}$ and $t_{90}$ are, respectively, the times when the total stellar mass reaches 0.1 and 0.9 times the final value. 
Figure \ref{zu0215_2} presents $R_{\rm n}$ at 
given metallicity for different cloud column densities. 
We here exclude the cases with the lowest $\Sigma < 10~{M_{\odot} \, {\rm pc^{-2}}}$ (open circles in Fig. \ref{zu3}) in evaluating the mean value as
those clouds are disrupted without forming the second sink particle.
We can see that $R_{\rm n}$ at given metallicity is almost constant regardless of the surface densities $\Sigma$.
We find that the mean cold gas fraction $\left< R_{\rm n} \right>$ decreases with decreasing metallicity as $\left< R_{\rm n} \right> = 0.67$, $0.36$ and $0.22$ at metallicity of $Z=1~Z_{\odot}$, $10^{-1}~Z_{\odot}$ and $10^{-2}~Z_{\odot}$, respectively. 

Figure \ref{zu_rhoT} shows the gas mass distribution on the number density and temperature plane for the cases with $(M_{\rm cl}, R_{\rm cl}) = (10^{5}~M_{\odot}, 20~{\rm pc})$ at $t=t_{\rm ff}$ ($4.7~{\rm Myr}$).
With $Z=1~Z_{\odot}$, dense star-forming gas with $n_{\rm H} \gtrsim 10^{4}~{\rm cm^{-3}}$ has low temperature of $30~{\rm K} < T < 200~{\rm K}$.
Most of this component is in the molecular state. 
At low metallicity of $Z=10^{-2}~Z_{\odot}$, on the other hand, there is few dense and cold gas with $T < 100$~K because of inefficient radiative cooling.
Star clusters directly form from such warm and atomic gases. 
The above features are also highlighted in Figure \ref{zu45.1}, showing the density-integrated temperature distribution for the same cases. 
Two peaks at 40-50K and 100K are observable in both cases corresponding to the different main cooling processes, the dust and [C{\sc ii}] cooling, respectively. 
The lower (higher) temperature peak is higher for 
$Z = Z_\odot$ ($Z = 10^{-2}~Z_\odot$, respectively).
At $Z=10^{-2}~Z_{\odot}$, more neutral gas distributes above $10^{2}~{\rm K}$ owing to less efficient [C{\sc ii}] cooling than in the solar metallicity case. In each case, plenty of gas accumulates around the peak temperature because the cooling becomes inefficient below that. Even if the gas is not affected by heating from the newborn stars, it stays at these temperatures for a long time. 
Thus, we consider the gas with $T < 200~{\rm K}$ as the star-forming gas in both cases.

\subsubsection{Star formation efficiencies}\label{SFEs_model}

Suppose a uniform density sphere whose mass and radius are $M_{\rm cl}$ and $R_{\rm cl}$.
Here we consider dynamics of the boundary shell of 
the HII region, which encloses the mass $M_{\rm sh} = M_{\rm cl} \left( r_{\rm sh} / R_{\rm cl} \right)^3$, where $r_{\rm sh}$ is the radius of the shell.
We calculate the motion of the shell driven by the thermal pressure as
\begin{eqnarray}
	\frac{d}{dt} \left( M_{\rm sh} \dot r_{\rm sh} \right) = 8 \pi k_{\rm B} T_{\rm i} n_{\rm i} r_{\rm sh}^2, \label{1228.1} 
\end{eqnarray}
where $T_{\rm i}$ is the temperature of ionized gas \citep[e.g.,][]{2002ApJ...566..302M, 2006ApJ...646..240H, 2009ApJ...703.1352K}.
The number density of the ionization region is given by the balance between the ionization and recombination rates as
\begin{eqnarray}
	n_{\rm i} = \left( \frac{3 f_{\rm ion} S_{\rm ion} }{4 \pi r_{\rm sh}^3 \alpha_{\rm B}} \right)^{1/2}, \label{1228.2} 
\end{eqnarray}
where $S_{\rm ion}$ and $\alpha_{\rm B}$ are the emissivity of ionization photons and the recombination rate coefficient.
The emissivity is proportional to the final total mass of sink particles as $S_{\rm ion} =  s_* \epsilon_{*} M_{\rm cl}$.
The photon production rate per unit mass 
$s_* = 8.0 \times 10^{46}~{\rm s^{-1}}~{M_{\odot}^{-1}}$, calculated from the same isochrone in Section \ref{radsource}.
The recombination rate coefficient is $\alpha_{\rm B}=2.6 \times 10^{-13} (T_{\rm i}/10^{4}~{\rm K})^{-0.8}~{\rm cm^3 \, s^{-1}}$\citep{2006agna.book.....O}.
In Equation \eqref{1228.2}, we introduce the factor $f_{\rm ion}$ that represents the fraction of ionizing photons used for ionization of hydrogen before they are absorbed by dust. 
We use the fitting function of $f_{\rm ion}$ obtained with the calculation of the hydrostatic structure inside H{\sc ii} regions \citep{2011ApJ...732..100D}.
More details of $f_{\rm ion}$ are described in Appendix \ref{fion}.
We introduce the following dimensionless parameters, 
\begin{eqnarray}
	x = r_{\rm sh} / R_{\rm cl}, \label{1228.3}
\end{eqnarray}
and 
\begin{eqnarray}
	\tau = \left[ \frac{48 \pi k_{\rm B}^2 T_{\rm i}^2 f_{\rm ion} s_* \epsilon_*}{\alpha_{\rm B} M_{\rm cl} R_{\rm cl}} \right]^{1/4} t. \label{1228.4}
\end{eqnarray}
With these new variables and equation \eqref{1228.2}, we rewrite equation \eqref{1228.1} as
\begin{eqnarray}
	\frac{d}{d \tau} \left( x^3 \frac{d}{d\tau} x \right) = x ^{1/2}. \label{1228.5}
\end{eqnarray}
Equation \eqref{1228.5} has the self-similar solution $x = \left( 7\tau/6 \right)^{4/7}$ \citep{2009ApJ...703.1352K}.
The ionization front reaches outside the cloud at $x=1$.
By setting $x=1$, we can derive the propagation time of the ionizing front 
\begin{eqnarray}
	t_{\rm HII} = \frac{6}{7} \left[ \frac{\alpha_{\rm B} M_{\rm cl} R_{\rm cl}}{48 \pi k_{\rm B}^2 T_{\rm i}^2 f_{\rm ion} s_{*} \epsilon_*} \right]^{1/4}. \label{eqtHII}
\end{eqnarray}
Here, we have assumed that the emissivity $S_{\rm ion}$ in Equation \eqref{1228.2} is given with the final stellar mass.
This simple approximation is appropriate because of the weak dependence of the propagation time $t_{\rm HII}$ on the emissivity $S_{\rm ion}$ (Eq. \ref{eqtHII}).

Assuming the star formation continues until $t \simeq t_{\rm HII}$, we can estimate the total stellar mass $M_*$ as
\begin{eqnarray}
	M_* = 	\epsilon_{*} M_{\rm cl} = \dot M_* t_{\rm HII}, \label{1228.8}
\end{eqnarray}
where $\dot M_{*}$ is the star formation rate.
As the cloud collapses, the gas accretes onto the central dense region at the rate $M_{\rm cl}/ t_{\rm ff}$ where $t_{\rm ff}$ is the free-fall time of the cloud. 
However, only a part of the accreted gas is converted into stars due to stellar feedback.  
Regarding the fraction of cold neutral gas $R_{\rm n}$ (Eq. \ref{EqRn}) as the conversion rate to the star clusters, we estimate the star formation rate as
\begin{eqnarray}
	\dot M_* = R_{\rm n} \frac{M_{\rm cl}}{t_{\rm ff}}, \label{0217.1}
\end{eqnarray}
where $t_{\rm ff} = \sqrt{3\pi/32 G \rho}$.
The star formation efficiency is estimated from Equation \eqref{eqtHII}, \eqref{1228.8} and \eqref{0217.1}, 
\begin{align}
	\epsilon_* 
	&= 0.18 ~ \left( \frac{R_{\rm n}}{0.67} \right)^{4/5} \left( \frac{\Sigma}{80~{M_{\odot} \, {\rm pc}^{-2}}} \right)^{1/2} \nonumber \\ 
	& \hspace{1cm} \left( \frac{M_{\rm cl}}{10^{5}~M_{\odot}} \right)^{1/10} \left( \frac{T_{\rm i}}{10^{4}~{\rm K}} \right)^{-14/25} \left( \frac{f_{\rm ion}}{0.3} \right)^{-1/5}, \label{1228.9} 	
\end{align}
where we use the relation between surface densities and radii of clouds as $\Sigma = M_{\rm cl}/ (\pi R_{\rm cl}^2)$.
In Figure \ref{zu3}, the shaded regions represent the SFEs at $M_{\rm cl} = 10^{4}$ and $10^{5}~M_{\odot}$.
Here, we use the temperature of ionized gas and the fraction of cold neutral gas $(T_{\rm i}, R_{\rm n})= (9\times10^3~{\rm K}, 0.67)$, $(1.6 \times 10^{4}~{\rm K}, 0.36)$ and $(2 \times 10^{4}~{\rm K}, 0.22)$ at $Z=1~Z_{\odot}$, $10^{-1}~Z_{\odot}$ and $10^{-2}~Z_{\odot}$, respectively, which are obtained from the simulations.
Also, $f_{\rm ion}$ is calculated from Equation \eqref{a2.1}.
Note in the dependency of SFE by Equation \eqref{1228.9}, 
the part of $\epsilon_* \propto M_{\rm cl}^{1/10} \Sigma^{1/2}$ depends on the initial cloud properties and that of 
$\epsilon_* \propto R_{\rm n}^{4/5}T_{\rm i}^{-14/25}f_{\rm ion}^{-1/5}$ on the metallicity. 
As can be seen in Figure \ref{zu3}, 
the semi-analytical model reproduces the simulation results well. 
At $Z=1~Z_{\odot}$, the analytical estimate for the model with the lowest-$\Sigma$ deviates from the numerical simulations because the conversion rate $R_{\rm n}$ decreases, as shown in Figure \ref{zu0215_2}.

Substituting Equation \eqref{1228.9} into \eqref{eqtHII}, we can obtain the expansion time of H{\sc ii} regions as 
\begin{align}
	t_{\rm HII} = 1.3 \times 10^{6} ~{\rm yr} &\left( \frac{R_{\rm n}}{0.67} \right)^{-1/5} \left( \frac{\Sigma}{80~M_{\odot}{\rm pc^{-2}}} \right)^{-1/4} \nonumber \\
	& \left( \frac{M_{\rm cl}}{10^5 ~M_{\odot}} \right)^{7/20} \left( \frac{T_{\rm i}}{10^{4}~{\rm K}} \right)^{-9/25} \left( \frac{f_{\rm ion}}{0.3} \right)^{-1/5}.  \label{0109}
\end{align}
The lifetime of cluster-forming clouds is the sum of the free-fall time and the expansion time of H{\sc ii} regions as $t_{\rm life} = t_{\rm ff} + t_{\rm HII}$.
In Figure \ref{zu335}, the solid and dashed lines show the lifetime of cluster-forming clouds estimated with Equation \eqref{0109}.
The analytical results coincide with the numerical ones.
Because of weaker dependence of $t_{\rm HII}$ on the surface density ($\propto \Sigma^{-1/4}$) compared with $t_{\rm ff}  (\propto \Sigma^{-1/2})$, the lifetime of cluster-forming clouds becomes significantly longer than $t_{\rm ff}$ as $\Sigma$ increases.

\subsection{Conditions of dust shielding}\label{optical_depth_of_shielding}

Turbulent motions create sheet-like density fluctuations as shown in Figures \ref{zu200108} and \ref{zu200108_2}. These structures survive even
after the H{\sc ii} region expansion
in the case of $Z=1~Z_{\odot}$  (Fig. \ref{zu200108}-3) while they completely disappear 
in the case of $Z=10^{-2}~Z_{\odot}$ (Fig. \ref{zu200108_2}-3). This difference comes from the fact that the optical depth of such structures against ionizing photons depends on metallicity as the dust grains are the main source of opacity. 
In the shade of optically-thick sheets, 
the gas can flow into the star-forming regions, and the star formation continues until the sheets are disrupted by the H{\sc ii}-region expansion.
Below we analytically estimate the optical depth of such sheets and clumps, and consider under what conditions such dense structures are protected against UV photons by dust attenuation. 
This is useful to understand the simulation results because these quantities should be related to the metallicity-dependence of the cold gas mass fraction $R_{\rm n}$ (Eq. \ref{EqRn}), which is higher with higher metallicity.

On large scales, where the turbulence is supersonic, the turbulent velocity field dominates the cloud evolution, and the gas 
accumulates into the sheet-like structures.
Star formation occurs after one free-fall time from the start of the collapse.
Until then, large-scale sheets are formed via the gas swept up on the scale where the crossing time $l/\sigma(l)$ is equal to the free fall time $t_{\rm ff}$:
\begin{eqnarray}
 l_{\rm c} = 7.4 ~{\rm pc} \left( \frac{R_{\rm cl}}{20~{\rm pc}} \right) \left( \frac{\alpha_0}{0.5} \right), \label{eq52}
\end{eqnarray}
where we use the scaling relation of velocity dispersion $\sigma(l) \propto l^{1/2}$ \citep{1981MNRAS.194..809L}, with the normalization given by $\sigma_0 = (3 \alpha_0 G M_{\rm cl}/ 5 R_{\rm cl})^{1/2}$ (see Eq. \ref{1.1}).
We estimate the optical depth of the sheet to ionizing photons with the cross section of dust grains $\sigma_{\rm d} = 10^{-21} ~{\rm cm^{2}} (Z/Z_{\odot})$ as
\begin{eqnarray}
	\tau_{\rm sh} = \sigma_{\rm d} n_{0} l_{\rm c} = 2.0 \left( \frac{\Sigma}{80~M_{\odot}\,{\rm pc^{-2}}} \right) \left( \frac{\alpha_0}{0.5} \right) \left( \frac{Z}{Z_{\odot}} \right), \label{eq54}
\end{eqnarray}
where $n_0$ is the mean number density of the cluster-forming cloud.
At $Z=1~Z_{\odot}$, the sheets are protected by photoionization if the surface density is higher than $80~M_{\odot} {\rm pc^{-2}}$.
In other cases, such as low-surface density or low-metallicity cases, dust attenuation becomes ineffective and the sheets are easily photoionized.

On the small scales, where the turbulent velocity becomes subsonic, the critical scale is estimated by comparing the turbulent velocity and sound speed \citep{2007ARA&A..45..565M}, 
\begin{eqnarray}
 l_{\rm s} = 0.4 ~{\rm pc} \left( \frac{\Sigma}{80~M_{\odot}\, {\rm pc^{-2}}} \right)^{-1} \left( \frac{\alpha_0}{0.5} \right)^{-1} \left( \frac{T}{20~{\rm K}} \right), \label{eq51}
\end{eqnarray}
where $T$ is the gas temperature.
At a scale smaller than $l_{\rm s}$, if the gas cools down and starts to collapse, the turbulent motion cannot suppress the collapse. 
Therefore, we consider $l_{\rm s}$ as the typical size of star-forming clumps. 
Given that $\lambda_{\rm J} = l_{\rm s}$, the number density becomes 
\begin{eqnarray}
	n_{\rm sf} = 1.9 \times 10^{4} ~{\rm cm^{-3}} \, \left( \frac{\Sigma}{80~M_{\odot}{\rm pc^{-2}}} \right)^{2} \left( \frac{\alpha_{0}}{0.5} \right)^{2} \left( \frac{T}{20~{\rm K}} \right)^{-1}. \label{nfl}
\end{eqnarray} 
Same as for Equation \eqref{eq54}, optical depth of the star-forming clouds $\tau_{\rm sf}$ is estimated with Equation \eqref{eq51} and \eqref{nfl} as
\begin{eqnarray}
	\tau_{\rm sf} = \sigma_{\rm d} n_{\rm sf} l_{\rm s} = 23 \left( \frac{\Sigma}{80~M_{\odot}{\rm pc^{-2}}} \right) \left( \frac{\alpha_{0}}{0.5} \right) \left( \frac{Z}{Z_{\odot}} \right). \label{tau_fl}
\end{eqnarray}
At $Z=1~Z_{\odot}$ the clump is shielded from UV radiation from neighboring stars.
If $Z \ll 0.1 ~Z_{\odot}$, the star-forming clouds are ionized, resulting in rapid quenching of star formation.

\section{Summary \& Discussion }\label{summary}
We have performed 3D RHD simulations to systematically study the star cluster formation and cloud dispersal in various environments, including very low-metallicity cases.
Our simulations include radiative feedback from massive stars, such as photoionization of hydrogen atoms, photoelectric heating, and photodissociation of molecules.
We have investigated the cluster-forming clouds with different surface densities and metallicities, $\Sigma = 10-300~M_{\odot}\,{\rm pc^{-2}}$ and $Z=10^{-2} - 1~ Z_{\odot}$. 
We have also developed a semi-analytical model that reproduces well the simulation results, e.g., the dependence of the star formation efficiency (SFE) on the cloud surface density and metallicity.
Our findings are summarized as follows:

\begin{description}
\item[(i)] 
Overall evolution in the simulation is qualitatively similar among all the examined cases. As a cloud collapses, the initial turbulent velocity field creates the dense gas filaments, which fragment to form stars.
As the mass in stars increases, UV radiation from newborn massive stars creates H{\sc ii} bubbles, 
which expands due to its high thermal pressure and eventually blow away surrounding non-ionized gas.
Once the radiative feedback becomes effective, the star formation is quenched within a free-fall time. That is, the natal cloud is promptly disrupted after the massive star formation.

\item[(ii)] 
At each metallicity, the SFE $\epsilon_*$ increases with increasing initial surface density $\Sigma$. At $Z=1~Z_{\odot}$, for instance, $\epsilon_* = 0.02$ for $\Sigma= 10~{\rm M_{\odot}\,{\rm pc^{-2}}}$ and $\epsilon_* = 0.3$ for $\Sigma= 300~{\rm M_{\odot}\,{\rm pc^{-2}}}$. 
At low metallicity, clusters form in atomic clouds as the molecule formation time is longer than the cooling or dynamical time. We have found that the SFE is systematically lower for lower metallicity with the reduction factor $\simeq 0.3$ at $Z = 10^{-2}~Z_\odot$, regardless of $\Sigma$. In low-metallicity environments, the temperature of the photoionized gas is high due to ineffective radiative cooling and dust attenuation of ionizing photons.
This leads to more rapid expansion of the H{\sc ii} regions, which terminates the star formation over the cloud in a shorter timescale. The newborn clusters are less gravitationally bound after the cloud dispersal than those at the solar metallicity.

\item[(iii)]
To understand the dependence of SFE on cloud properties and metallicity, we have developed a semi-analytical model that gives SFE agreeing 
excellently with the simulation results. 

\end{description}

CO molecules are the standard tracer of the molecular clouds, i.e., the star-forming cold gas. Since we have consistently solved the thermal and chemical states of the atomic and molecular gas in our simulations, we could follow the time evolution of the CO-to-H$_2$ conversion factor of the cluster-forming clouds \citep[e.g.,][]{Inoguchi20}, although not focused above. In our simulations for $Z=1~Z_{\odot}$, the formation of CO and $\rm H_{2}$ molecules occurs in the dense filaments, in which newborn clusters are formed. 
In the cases of $Z=10^{-2}~Z_{\odot}$, molecule formation hardly proceeds and the star formation occurs directly from almost atomic gas \citep{2012ApJ...759....9K}. Observations suggest a high value of the CO-to-${\rm H_2}$ conversion factor for GMCs with sub-solar metallicity associated with massive star-forming regions \citep{2009ApJ...702..352L}. Our simulations should potentially provide such metallicity dependence even for very metal-poor cases with $Z = 10^{-2}~Z_\odot$, which will be considered in our future studies.

In our simulations, we only considered radiative feedback and did not include kinematic feedback effects, such as outflows, stellar winds, and supernovae. 
Outflows alone would not suppress star formation in star-forming clouds whose escape velocities are typically larger than $1~{\rm km\, s^{-1}}$ \citep{2015ApJ...815...68M, 2019ARA&A..57..227K}. 
However, outflows inject momentum to the surrounding medium and maintain turbulent motions in cluster-forming clouds \citep{2006ApJ...640L.187L, 2007ApJ...662..395N}. This is likely to suppress the SFE.
Stellar winds can also push the neutral gas away.
The star formation efficiencies may further be reduced due to the combined effects of stellar winds and photoionization \citep[e.g.,][]{2014MNRAS.442..694D}.
When the duration of star formation in clouds is longer than the lifetime of OB stars, supernovae also contribute to destruction of the clouds \citep[e.g.,][]{2016MNRAS.463.3129G}.
In future works, we will take those feedback effects into account in our numerical simulations.

Our simulations show that the cluster formation and cloud dispersal proceed in the filamentary gas. 
Given that a strong magnetic field exists in the cloud, the gas structure will change significantly \citep[e.g.,][]{2012ApJ...759...35I, 2015A&A...580A..49I}. 
Also, the magnetic fields give support to collapsing GMCs and delay the star formation \citep{2015MNRAS.454.4484G}.
We plan to study the impact of magnetic fields combined with the radiative feedback on star formation in GMCs in future work.

\section*{Acknowledgements}
The authors wish to express their cordial thanks to Profs. Masayuki Umemura and Ken Ohsuga for their continual interest, advice and encouragement. 
This work is supported by the Grants-in-Aid for Basic Research by the Ministry of Education, Science and Culture of Japan (HY: 17H04827, 18H04570, 20H04727, TH: 19H01934, KO: 25287040, 17H01102, 17H02869) and NAOJ ALMA Scientific Research Grant Numbers 2019-11A. 
The numerical simulations were performed on the Cray XC50 (Aterui II) at the Center for Computational Astrophysics (CfCA) of National Astronomical Observatory of Japan and the Cray XC40 at Yukawa Institute for Theoretical Physics in Kyoto University

\section*{DATA AVAILABILITY}
The data underlying this article will be shared on reasonable request to the corresponding author.


\bibliographystyle{mnras}
\input{./paper4.bbl}

\appendix
\section{Radiative process}\label{apdA}
As discussed in Section \ref{chemical_thermal}, we use the adoptive ray-tracing method for radiation transfer \citep{2002MNRAS.330L..53A}.
In this method, we calculate the equation of radiation transfer from each radiation source, discretizing the solid angle with HEALPix \citep{2005ApJ...622..759G}.
In the regions far from the radiation sources, we split the ray to ensure that three rays always cross each cell.

In raytracing, we calculate the column densities of atomic hydrogen $N_{\rm HI}$, hydrogen atom $N_{\rm Hn}$, $\rm H_2$ and CO ($N_{\rm H_2}$, $N_{\rm CO}$).
The EUV radiation energy flux is calculated as 
\begin{eqnarray}
	F_{\rm EUV} = \frac{ L_{\rm *, EUV}}{4 \pi r ^2} \exp[- ( \sigma_{\rm HI} N_{\rm HI} + \sigma_{\rm d} N_{\rm Hn} )], \label{a.3.1} 
\end{eqnarray}
where $L_{\rm *, EUV}$ is the EUV luminosity of each radiation source.
To obtain the cross-sections of atomic hydrogen and dust grains, we pre-calculate the frequency averaged cross-section with the energy spectrum of radiation sources in the frequency range $13.6~{\rm eV} \leqq h \nu \leqq 10^3 ~{\rm eV}$.
We use the approximation formula of the cross-section for hydrogen atoms as 
$\sigma_{\rm HI} (\nu) = 6.3 \times 10^{-18} \left( \nu / \nu_l \right)^{-3} ~{\rm cm^2}$ \citep{2006agna.book.....O} where $\nu_l$ is the Lyman limit frequency.
The cross-sections of dust grains are given by \citet{1993ApJ...402..441L}, assuming that the dust-to-gas mass ratio is 0.01 at $Z=1~Z_{\odot}$.
We consider only the dust attenuation for the FUV light and the continuum components as 
\begin{eqnarray}
	F_{\rm i} = \frac{ L_{\rm *, i}}{4 \pi r ^2} \exp[- \sigma_{\rm d, i} N_{\rm Hn} ] \hspace{1mm} ({\rm i = FUV, continuum}), \label{a.3.3} 
\end{eqnarray}
where $\sigma_{\rm i}$ is the frequency averaged cross section in the frequency range $11.2 {\rm eV} \leqq h\nu \leqq 13.6 ~{\rm eV}$ ($ h \nu \leqq 11.2 {\rm eV}$) for FUV photons (continuum components, respectively).
We include the radiation force caused by absorption on dust grains and neutral hydrogen at the three components of the radiation field in the hydrodynamics calculations.
Note that, in our simulations, the radiation force is secondary compared to the photo-ionization heating.

In chemical and thermal solvers, we consider photoionization (H{\sc i} and O{\sc ii}), photodissociation ($\rm H_{2}$, and CO), photoelectric heating and dust heating as follows.

The photoionization rates for H{\sc i} and O{\sc ii} is obtained as
\begin{eqnarray}
	R_{\rm HI} = \int ^{\infty}_{\nu _{l}} d \nu \sigma_{\rm HI} (\nu) \frac{F_{\nu}}{h \nu}, \label{a.3.3.2}
\end{eqnarray}
\begin{eqnarray}
	R_{\rm OII} = \int ^{\infty}_{\nu _{\rm OII}} d \nu \sigma_{\rm OII} (\nu) \frac{F_{\nu}}{h \nu}, \label{a.3.4}
\end{eqnarray}
where
$\nu_{l}$ and $\nu_{\rm OII}$ are the ionization limit frequency for H{\sc i} and O{\sc ii}.
Here, we use the cross section of H{\sc i} ($\sigma _{\rm HI}$) and O{\sc ii} ($\sigma _{\rm OII}$) of \citet{2006agna.book.....O}.

We use the $\rm H_2$ photodissociation rate of \citet{1996ApJ...468..269D} as \citet{2020arXiv200402364F}, considering dust absorption and self-shielding. 
We estimate the photodissociation rate of CO and the photoelectric heating rate with the same method of \citet{2018ApJ...857...57N}.
The photodissociation rate of CO molecules is given as \citep{1996A&A...311..690L}
\begin{eqnarray}
	R_{\rm CO} = G_{\rm FUV} p_{\rm diss} n_{\rm CO} \Theta_1(N_{\rm CO}) \Theta_2 (N_{\rm H_2}) \Theta_3 (A_{\rm V}), \label{a1.1} 
\end{eqnarray}
where $n_{\rm CO}$, $A_{\rm V}=5.3 \times 10^{-22} N_{\rm Hn} (Z / Z_{\odot})$ and $p_{\rm diss} = 1.03 \times 10^{-10}~{\rm s^{-1}}$ are the number density of CO, the visual extinction and the unshielded CO photodissociation rate.
The strength of FUV is represented with the averaged interstellar flux $F_{\rm ISRF} = 1.6 \times 10^{-3}~{\rm erg \, cm^{-1} \, s^{-1}}$ as $G_{\rm FUV} = L_{\rm FUV}/ ( 4 \pi r^2 F_{\rm ISRF} )$ where $L_{\rm FUV}$ is the FUV luminosity.
The rates of self-shielding , $\rm H_2$ shielding, and dust shielding ($\Theta_1$, $\Theta_2$ and $\Theta_3$) are given in Tables of \citet{1996A&A...311..690L}.

We use the analytical formula of \citet{1994ApJ...427..822B} for the photoelectric heating as
\begin{eqnarray}
	\Gamma_{\rm pe} = 10^{-24} \epsilon_{\rm pe} n_{\rm H} G_{\rm FUV} \exp(- 1.8 A_{\rm V}) \left( \frac{Z}{Z_{\odot}} \right), \label{a1.2}
\end{eqnarray}
where the exponential factor represents the dust absorption \citep{2018ApJ...857...57N}, and $\epsilon_{\rm pe}$ is given with $G_{\rm FUV}$ and the number density of electron as
\begin{eqnarray}
	\epsilon_{\rm pe} = \frac{4.87 \times 10^{-2}}{1+4\times10^{-3}\gamma_{\rm pe}^{0.73}} + \frac{3.65 \times 10^{-2} (T/10^4)^{0.7}}{1 + 2\times 10^{-4} \gamma_{\rm pe}}, \label{a1.3} 
\end{eqnarray} 
where $\gamma_{\rm pe} = G_{\rm FUV} \exp(- 1.8 A_{\rm V}) \sqrt{T} / n_{\rm e}$.

To obtain the dust temperature, we calculate the balance among the dust thermal emission, absorption of the direct light (EUV, FUV and continuum), and energy transport from gases.
We use the energy transport rate between dust grains and gases given by \citet{1979ApJS...41..555H}.
Further details are shown in \citet{2020arXiv200402364F}.

\section{Hydrogen absorption fraction of ionizing photons}\label{fion}

In Section \ref{SFEs_model}, we use the fitting formula of \citet{2011ApJ...732..100D} for a fraction of EUV photons absorbed by hydrogen atoms in H{\sc ii} regions.
\citet{2011ApJ...732..100D} construct the model of the hydrostatic structure of H{\sc ii} regions, including the thermal pressure and radiation pressure induced by photoionization and dust grains.
When radiation pressure becomes strong enough, the shell structure is formed inside the H{\sc ii} regions and the recombination of hydrogen is enhanced due to the high density of the shell.
We incorporate these dynamical effects for $f_{\rm ion}$, although the constant density is assumed in H{\sc ii} regions in our semi-analytical model.
The fraction of hydrogen ionization for EUV photons $f_{\rm ion}$ is given as 
\begin{eqnarray}
	f_{\rm ion} = \frac{1}{1 + (2/3 + AB) \tau_{\rm d,0}} + \frac{AB \tau_{\rm d,0}}{1 + B \tau_{\rm d,0}}, \label{a2.1}
\end{eqnarray}
where $\tau_{\rm d,0}$ is the optical depth of H{\sc ii} region as 
\begin{eqnarray}
	 \tau_{\rm d,0} = 2.1 \left( \frac{\sigma_{\rm d}}{10^{-21} ~{\rm cm^{-2}}} \right) \left[ \left( \frac{S_{\rm ion}}{10^{49}~{\rm s^{-1}}} \right) \left( \frac{n_{\rm i}}{10^{3}~{\rm cm^{-3}}} \right) \right]^{1/3} \left( \frac{T_{\rm i}}{10^{4}~{\rm K}} \right)^{0.28}. \label{a2.2}
\end{eqnarray}
In Equation \eqref{a2.2}, $S_{\rm ion}$, $n_{\rm i}$ and $T_{\rm i}$ are the emissivity of ionizing photons, the number density and the temperature in the H{\sc ii} regions.
Here, we use the cross section of dust grains as $\sigma_{\rm d} = 10^{-21} ~{\rm cm^{-2}} (Z/Z_{\odot})$.
In the original formula of \citet{2011ApJ...732..100D}, they use the root mean square of the number density in Equation \eqref{a2.2}. 
It is not significantly different from the uniform density $n_{\rm i}$ in our parameter setups.
The parameters of $A$ and $B$ are given as
\begin{eqnarray}
	A = \frac{1}{1 + 0.75 \gamma^{0.65} \beta^{-0.44}}, \label{a2.3}
\end{eqnarray}
\begin{eqnarray}
	B = \frac{0.5}{1 + 0.1 (\gamma /\beta)^{1.5}}, \label{a2.4}
\end{eqnarray}
where $\beta$ and $\gamma$ are determined by the luminosity ratio of ionizing and non-ionizing radiation ($L_{\rm n}$, $L_{\rm i}$) and the temperature of H{\sc ii} regions as 
\begin{eqnarray}
	\beta = \frac{L_{\rm n}}{L_{\rm i}}, \label{a2.5}
\end{eqnarray}
\begin{eqnarray}
	\gamma = 11.2 \left(\frac{T_{\rm i}}{10^{4} ~{\rm K}}\right)^{1.83} \left( \frac{18~{\rm eV}}{<h \nu>} \right) \left( \frac{\sigma_{\rm d}}{10^{-21} ~{\rm cm^{-2}}} \right) , \label{a2.6}
\end{eqnarray}
In Equation \eqref{a2.6}, we assume that the average energy of ionizing photons becomes $<h \nu>=18~{\rm eV}$, which is the typical one for the stellar IMF, metallicity and age in our simulations.

\bsp	
\label{lastpage}
\end{document}